\documentclass{aa}  
\usepackage{graphicx} 
\usepackage{xcolor} 
\usepackage{txfonts}
\usepackage{array}
\usepackage{subfigure}
\usepackage[export]{adjustbox}
\usepackage{multirow,tabularx}

\newcommand{\cathode}{C\textsc{athode}}
\newcommand{\anode}{A\textsc{node}}

\newcommand{\vm}{\textsc{Via}~\textsc{Machinae}}

\newcommand{\vma}{VM3--A}
\newcommand{\vmc}{VM3--C}

\AddToHook{begindocument/before}{\usepackage{hyperref}}

\begin{document}

\title{Via Machinae 3.0: A search for stellar streams in \textit{Gaia} with the CATHODE algorithm}


   \author{Anna Hallin\inst{1}
          \and
          David Shih\inst{2}
          \and
          Claudius Krause\inst{3}
          \and
          Matthew R.~Buckley\inst{2}
          }
\institute{Institut für Experimentalphysik, Universit\"at Hamburg, Luruper Chaussee 149, 22761 Hamburg, Deutschland \\ \email{anna.hallin@uni-hamburg.de}
        \and
        NHETC, Department of Physics and Astronomy, Rutgers University, Piscataway, NJ 08854, USA  
    \and
        Marietta-Blau-Institut für Teilchenphysik (MBI Vienna), Österreichische Akademie der Wissenschaften (ÖAW), Dominikanerbastei 16, A-1010 Wien, Österreich }
        \date{\today}

\abstract{
   We apply the model-agnostic anomaly detection method \cathode{} --- originally developed for particle physics --- to search for stellar streams in \textit{Gaia} data. We combine \cathode{} with \vm{} 3.0: a re-optimized version of the stellar stream detection method that was 
   previously applied to {\it Gaia} data together with the related anomaly detection technique \anode{}. We demonstrate that the combination of \vm{} 3.0 with \cathode{}, called \vmc, not only re-discovers previously known streams, but also confirms many candidate streams identified in combination with \anode{} (denoted \vma). Compared to \vma{}, the number of stream candidates detected by \vmc{} increases by around 10\%. Moreover, both of the methods discover the same two large clusters of stream candidates in the Northern Galactic hemisphere. We dub these highly significant anomalous structures the {\it Raritan} stream and the {\it Passaic} stream. These two structures
   may indicate the presence of larger objects, such as dwarf galaxy streams, or non-trivial orbital dynamics resulting in bifurcation or fanning, and are promising and high-priority targets for further analysis.
}

\maketitle

\section{Introduction}

The tidal stripping of stars from dwarf galaxies and globular clusters forms elongated, kinematically cold stellar streams. These unique systems can act as probes of the formation history \citep{1998ApJ...495..297J,1999MNRAS.307..495H,2006ApJ...642L.137B,2018ApJ...862..114S,2018MNRAS.478..611B}, gravitational potential \citep{1999ASPC..194...15J,2001ApJ...551..294I,2010ApJ...712..260K,2010DDA....41.0501N,2011MNRAS.417..198V,2012JCAP...08..027P,2013MNRAS.433.1813S,2015ApJ...799...28P,2014ApJ...795...94B,2015ApJ...803...80K,2015ApJ...801...98S,2016ApJ...833...31B,2019MNRAS.486.2995M,2019ApJ...883...27N,2021MNRAS.502.4170R,2022arXiv221014983P,2024ApJ...967...89I,2025arXiv250802666N} and dark matter substructure of the Milky Way \citep{2012ApJ...760...75C,2016MNRAS.457.3817S,2017MNRAS.470...60E,2019ApJ...880...38B,2019MNRAS.484.2009B,2020ApJ...892L..37B,2022MNRAS.513.3682D}. 

In the era of the {\it Gaia} space telescope,
stellar stream astronomy has seen a rapid growth in techniques and discoveries~\citep{Bonaca:2024dgc}. In particular, motivated by the size of the \textit{Gaia} dataset, there has been significant emphasis on automated stream discovery. A major effort was initiated by \textsc{Streamfinder}  \citep{10.1093/mnras/sty912} who searched for greater-than-expected numbers of stars along orbits and isochrones. While this assumes a model for the Milky Way potential and for stellar populations within the streams, it has proven to be highly successful, resulting in many new streams discovered in the \textit{Gaia} data \citep{2018MNRAS.478.3862M,2019ApJ...872..152I,2020ApJ...891L..19I,2021ApJ...914..123I,2021ApJ...920...51M,2024ApJ...967...89I}. Other searches for stellar streams within the \textit{Gaia} data include the works of \cite{2018ApJ...863...26Y,2019A&A...621L...3M,2020MNRAS.492.1370B} and \cite{2019A&A...622L..13M}. 

At the same time, there has been significant development of model-agnostic search strategies for stellar streams
~\citep{Shih:2021kbt,Shih:2023jfv,Pettee:2023zra,Sengupta:2024ezl} powered by modern machine learning (ML).  This was initiated in the \vm{} works of~\cite{Shih:2021kbt,Shih:2023jfv}, where it was realized 
that the search for stellar streams in Gaia kinematic data could be thought of as a type of {\it resonant anomaly detection}.
First motivated in the context of new physics searches at the Large Hadron Collider (LHC) (see 
\cite{Kasieczka:2021xcg,Aarrestad:2021oeb,Karagiorgi:2022qnh,Belis:2023mqs} for reviews), resonant anomaly detection is built on the idea of a ``bump hunt": a search for 
anomalies localized in a single {\it resonant feature} $m$. One defines {\it signal regions} (SRs) as windows in the resonant feature, $m_1<m<m_2$. Generally one uses the {\it sideband} (SB) regions $m<m_1$ and $m>m_2$ to model the backgrounds in the SR in a data-driven way. Resonant anomaly detection goes beyond simple, traditional one-dimensional bump hunts by leveraging modern ML methods to search in additional features $x$ where the anomalies may also be localized. In a local patch of the sky, stellar streams should be localized in both proper motions and should look line-like in angular coordinates. In principle they should also be localized to an isochrone in color and magnitude.  By searching for correlated overdensities in all of these features, resonant anomaly detection can  be sensitive to a wide-range of stellar streams and potentially other interesting coherent stellar substructures.

 The core of the original \vm{} works was the \anode{} algorithm~\citep{Nachman:2020lpy}. The aim of \anode{} is to learn an ``anomaly score'' $R(x)$ from the data which characterizes how overdense a point $x$ in the data is relative to the smooth background. \anode{} accomplishes this by training density estimators, in the form of normalizing flows (for reviews and original references, see \cite{2019arXiv190809257K,2019arXiv191202762P}), on data in the SR and SB. In~\cite{Shih:2021kbt}, \vm{} was first applied as a proof of concept to re-discover the extremely cold and bright stream GD-1 \citep{Grillmair_2006}. Then in \cite{Shih:2023jfv} --- henceforth referred to as VM2 --- it was applied to all of {\it Gaia} Data Release 2 (DR2) \citep{Gaia_DR2_2018} to perform an all-sky search for stellar streams. After selecting the most anomalous stars using the anomaly score, VM2 includes a number of steps to home in on stream-like overdensities. These include line-finding with the Hough transform,\footnote{For another application of the Hough transform to stellar stream searches in  extragalactic data, see \cite{2022ApJ...926..166P}.} clustering in proper motion space, and linking ``protostreams'' across multiple patches of the sky into full-fledged stream candidates. The end  result of the VM2 all-sky scan was the discovery of over 100 stream candidates with an estimated false positive rate of $\sim 10\%$.

In this paper, we present the first application of  the \cathode{}~\citep{Hallin:2021wme} resonant anomaly detection algorithm to the search for stellar streams. \cathode{} is another technique for learning the overdensity anomaly score $R(x)$ that combines normalizing flows with weak supervision (classifiers between data and background samples). While closely related, \cathode{} was shown to outperform \anode{} in the context of LHC physics in terms of its sensitivity to unknown anomalies. Here we confirm the superior performance of \cathode{}, first with GD-1, where it shows a much improved ability to separate signal from background compared to \anode{}; and then with an all-sky scan of {\it Gaia} DR2, where \cathode{} discovers $\sim 10\%$ more high-confidence stream candidates than previously found with \anode{}.

Since the main change from VM2 in this work is the replacement of \anode{} with \cathode{} in the estimation of the anomaly score, we will refer to this new stream search as 
\vm\ 3.0 --- \cathode{} Edition, or \vmc{} for short. While the changes we make to the post-anomaly detection steps in the stream-finding algorithm are minor, they are enough to slightly change the catalog of streams found previously with \anode{} in VM2. We will therefore also produce a new catalog of stream candidates using the \anode{} anomaly score, which will be referred to as \vma{} to distinguish it from VM2.

Despite the improved performance of \cathode{} over \anode{}, we consider the \vmc{} and \vma{} stream candidate catalogs to be complementary in many respects.
The main boost from \cathode{} appears to be for streams that have large signal-to-background ratios, consistent with what was found in the original work in the LHC context~\citep{Hallin:2021wme,Golling:2023yjq}. For lower signal to noise streams, both \cathode{} and \anode{} provide noisy estimates of the anomaly score, and so the two methods can be viewed as semi-independent stream searches. Candidates found by both methods are more likely to be real, and the union of both \vmc{} and \vma{} is a more comprehensive catalog of potential stream candidates.

Besides GD-1, \vmc{} and \vma{} both confirm more than ten previously-discovered stream candidates.
Some notable highlights here include Lethe~\citep{Grillmair_2009}, where to our knowledge \vm{} is providing the first ever measurements of the proper motions of its stars; Gaia-6, where \vma{} may be discovering an extension beyond what was reported in \citep{2019ApJ...872..152I}; and Gaia-1 \citep{2018MNRAS.481.3442M} and Jhelum \citep{2018ApJ...862..114S}, where \vmc{} confirms the extensions that were previously discovered in VM2. 

Beyond previously-known streams, \vmc{} and \vma{} together discover 93 new stream candidates, including 17 ``gold plated'' candidates that are discovered by both methods. Especially notable among these are two extremely complex clusters of stream candidates located in the northern Galactic hemisphere that we are dubbing the {\it Raritan} stream and the {\it Passaic} stream. More complicated than the typical globular cluster streams that \vm{} is designed to find, these are large and more diffuse kinematic structures that could be fragments of a broader dwarf galaxy stream (we see something analogous with \vm{} for the Sagittarius stream), or could be a sign of fanning or bifurcation due to nontrivial interactions between the orbital path and the Galactic potential  \citep{2007ApJ...659.1212M,2008ApJ...689..936J,2015MNRAS.452..301F,2015MNRAS.446.3100H,2015ApJ...799...28P,2016MNRAS.455.1079P,2020MNRAS.492.4398M,2021MNRAS.501.1791Y,2023ApJ...954..215Y,2025ApJ...979...75G}. These stream clusters, seen by both \vmc{} and \vma{}, are possibly the most interesting new discoveries found by \vm{} and are the highest-priority targets for future follow-up study.

This paper is organized as follows: Section~\ref{sec:dataset} describes the dataset. In Section \ref{sec:resonant_anomaly_detection} the resonant anomaly detection method with \anode{} and \cathode{} is described. Section \ref{sec:method} describes the \vm{} stream-finding algorithm. Our results are presented in Section~\ref{sec:results}, with a discussion and conclusion in Section~\ref{sec:discussion_conclusions}.

\section{Dataset}
\label{sec:dataset}
\subsection{The \textit{Gaia} space telescope}
\label{subsec:Gaia}
The \textit{Gaia} space telescope \citep{Gaia2016} was launched in 2013 by the European Space Agency (ESA). It is capable of measuring the angular position, parallax, and proper motion of bright stars (magnitude $\sim 20$) across the entire sky with nearly uniform coverage. In addition, the \textit{Gaia} spectrometer is capable of measuring the radial velocity of stars brighter than magnitude $\sim 14$, though we will not use this radial information in the current analysis. 

In this work we use \textit{Gaia} Data Release 2 (DR2) \citep{Gaia_DR2_2018} rather than DR3 \citep{Gaia_EDR3_2021,Gaia_DR3_2022},  so that our results can be directly comparable to ~\cite{Shih:2021kbt,Shih:2023jfv}.
DR2 is based on 22 months of data collection. Of the 1.7 billion objects in this dataset, about 1.3 billion stars have measured locations on the sky, parallax and proper motion in 2D. 

Following~\cite{Shih:2021kbt,Shih:2023jfv}, the stars in the \textit{Gaia} sky are divided into overlapping circular patches with a radius of $15^{\circ}$. In order to prevent distortion near the celestial poles, the angular position ($\alpha,\delta$) and corresponding proper motions ($\mu_{\alpha}*, \mu_{\delta}$) are transformed (using the \textsc{AstroPy} \citep{astropy:2013,astropy:2018,astropy:2022} coordinate transformation algorithm) to a latitude and longitude centered in the particular patch. The new angular position coordinates are $\phi$ (longitude) and $\lambda$ (latitude), with the corresponding proper motion coordinates $\mu_{\phi}\cos\lambda\equiv\mu_{\phi}^*$ and $\mu_{\lambda}$. The magnitude $g$ is used as-is, and color is defined as the difference between the blue and red passbands, $(b-r)$. The final variables for our analysis are then $(\phi,\lambda,\mu_{\phi}^*,\mu_{\lambda},g,b-r)$.

\begin{figure*}
    \centering
    \includegraphics[trim={0 2.5cm 3cm 2cm},clip,width=0.75\textwidth]{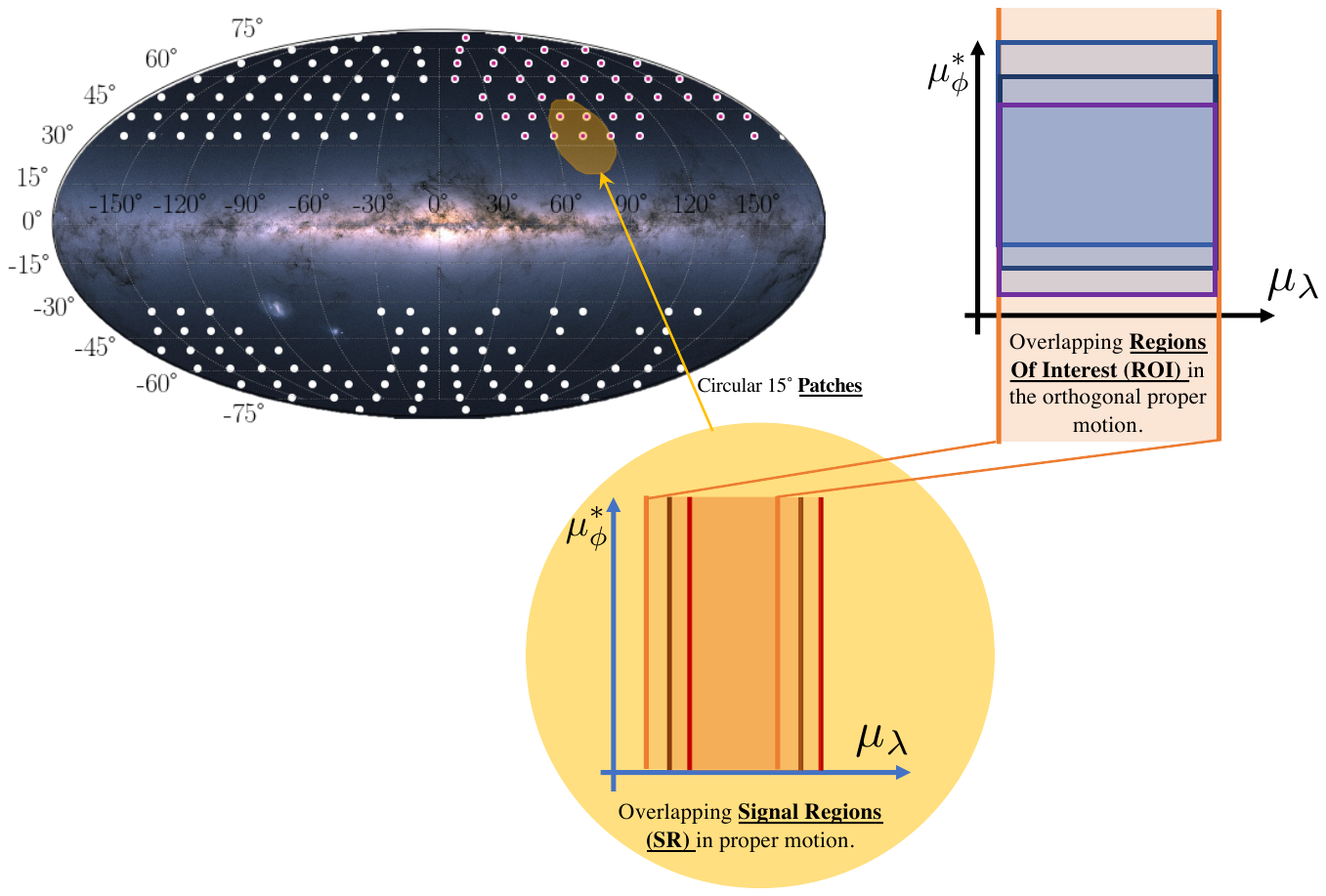}
    \caption{The sky is divided into patches, which are then divided into overlapping signal regions based on one of the proper motions, which are subdivided further into overlapping regions of interest along the orthogonal proper motion. The Galactic disk, the Large Magellanic Cloud and patches with high amounts of dust have been excluded from the all-sky scan. The centers of the remaining patches are indicated by the circles in the sky map. The filled-in circles in the upper right quadrant correspond to the 44 patches in the \textit{Galaxia} mock stellar catalog. Figure reproduced from \cite{Shih:2023jfv}.  }
    \label{fig:regions}
\end{figure*}

We exclude patches that are too close to the Galactic disk, as well as the patches containing the Large Magellanic Cloud and patches with high amounts of dust. This leaves a total of 163 patches that form the basis of our \textit{Gaia} DR2 stream search. The location of the centers of the patches are shown in Figure~\ref{fig:regions}.

\subsection{\textit{Galaxia} mock stellar catalog}
\label{subsec:Galaxia}

The \textit{Gaia} data does not include labels indicating whether a star belongs to a stellar stream or not, and so a method is needed to estimate the fraction of false positives among the stream candidates. To this end, we use a second dataset \citep{Sharma_2011,Rybizki_2018}, which we will refer to as \textit{Galaxia}. This is a simulated dataset, which contains distributions of the Milky Way halo, thick disk and bulge. These distributions are all smooth and without substructure; in particular, they contain no stellar streams. By comparing the \textit{Galaxia} result to the \textit{Gaia} result, it is possible to estimate the fraction of false positives among the stream candidates from both \cathode{} and \anode{}. In order to save computing time, only a quarter of the sky (44 patches out of 163 for the whole sky) is compared between \textit{Gaia} and \textit{Galaxia}. 

\section{Resonant anomaly detection with \anode{} and \cathode{}}
\label{sec:resonant_anomaly_detection}

\begin{figure}[t!]
    \centering
    \framebox{\includegraphics[width=0.475\textwidth]{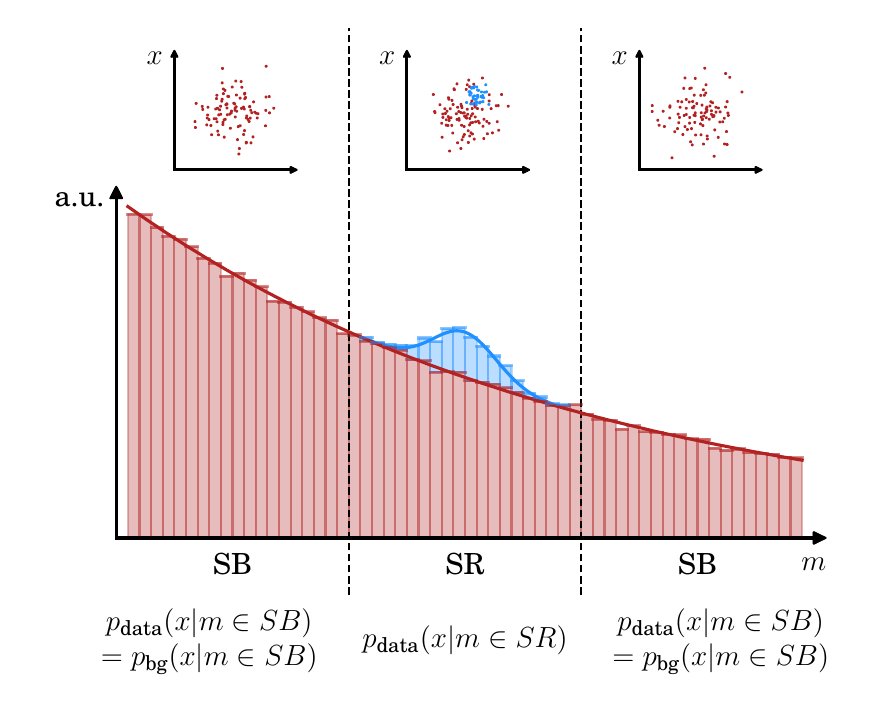}}
    \caption{Schematic view of resonant anomaly detection, reproduced from~\cite{Hallin:2021wme}. The signal (blue) is localized in the signal region (SR). The background (red) is estimated from a sideband region (SB).}
    \label{fig:SB-SR}
\end{figure}

In resonant anomaly detection, one first divides up the data features into a {\it resonant feature} $m$ where the signal is assumed to be localized (here taken to be one of the proper motions), and additional {\it anomaly features} $x$ (here taken to be the angular positions, color, magnitude, and the orthogonal proper motion). The signal is also expected to be localized in some (or all) of the anomaly features. 
The resonant feature $m$ is used to define intervals called signal regions (SRs); the complementary regions are called sidebands (SBs). Figure~\ref{fig:SB-SR} shows a sketch of the principle. The general strategy of
resonant anomaly detection is to use the data in the SB and SR to learn some kind of anomaly score $R(x)$, and to cut on this score in the SR to reveal any potential anomalies hiding in the data.

\anode{}~\citep{Nachman:2020lpy} and \cathode{}~\citep{Hallin:2021wme} are closely related resonant anomaly detection methods originally motivated by the search for new physics at the LHC.\footnote{\cathode{} has recently been applied by the CMS collaboration \citep{CMS:2024nsz} to the full Run~2 dataset.} In both, the aim is to estimate the likelihood ratio between data and background in the SR
\begin{equation}
\label{eq:Rofx}
R(x)=\frac{{p_{\mathrm{data}}(x)}}{ p_{\mathrm{bg}}(x)}.
\end{equation}
This is the optimal test statistic for detecting overdensities in the data relative to the background~\citep{neyman1933ix} and is commonly referred to as the ``optimal anomaly score''.

In \cathode{}, there are three steps in the estimation of this anomaly score:

\begin{enumerate}
    \item Train a conditional normalizing flow \citep{NIPS2017_6c1da886} to learn $p_{\rm data}(x|m)$ in the SB region. By assumption the SB is signal-poor, so $p_{\rm data}(x|m\in \mathrm{SB})=p_{\rm bg}(x|m\in \mathrm{SB})$ to a good approximation. For a narrow enough SR, the conditional normalizing flow can be expected to smoothly interpolate into the SR, yielding a model for the background density in the SR, $p_{\rm bg}(x|m\in \mathrm{SR})$. 
    
    \item Sample from the learned flow model for $p_{\rm bg}(x|m\in \mathrm{SR})$ to obtain a collection of events following the estimated background density in the SR. 
    
    \item Train a binary classifier to distinguish the sampled events from the actual events in the SR. With a sufficiently accurate background model, a well-trained classifier should approximate the data-to-background likelihood ratio $R(x)$. 
\end{enumerate}
In \anode{}, the first step also consists of training a conditional flow in the SB region. However, instead of sampling and subsequently training a classifier, \anode{} trains a second, independent normalizing flow in the SR and the anomaly score $R(x)$ is formed directly by taking the ratio of the two densities. As density estimation is generally a more difficult ML task than binary classification, \cathode{} is generally expected to be more sensitive to anomalies than \anode{}~\citep{Hallin:2021wme,Golling:2023yjq}. 

The single most computationally expensive step in this method is training the density estimator in step 1. Since this step is shared with \anode{}, we were able to re-use the sideband density estimators trained in VM2 for \vmc{} and \vma{}. While this saved significant computational resources, there is an associated cost: as we reuse this aspect of the VM2 calculation, the stream candidates found by \vmc{} are not fully independent of the previous results. We cannot rule out the possibility that some false positives in \vmc{} are due to systematic or statistical errors in the background models common to both analyses.

In order to deal with the limited dataset that is available here, \cathode{} was adjusted to use $k$-fold cross validation\footnote{$k$-folding divides the data into $k$ parts and trains the model $k$ times, each time with a different part of the data held out as test set. This allows the the model train on the entire dataset while still withholding some data for testing.} to make the most of the data. Based on test runs on a stellar catalog~\citep{Price_Whelan_2018} of the bright and well-characterized stream GD-1~\citep{Grillmair_2006}, four $k$-folds were deemed satisfactory. A hyperparameter scan for the classifier run on the GD-1 data showed that the optimal performance occurred when using two hidden layers with 512 nodes each and a ReLu as activation function. The classifier was trained for 75 epochs, using binary cross-entropy as the loss and Adam~\citep{kingma2017adam} as optimizer with a learning rate of 0.001. The ten epochs with the lowest validation loss were selected, and their predictions averaged in order to calculate the $R$ value.

\begin{figure*}
\centering
\includegraphics[width=0.75\textwidth]{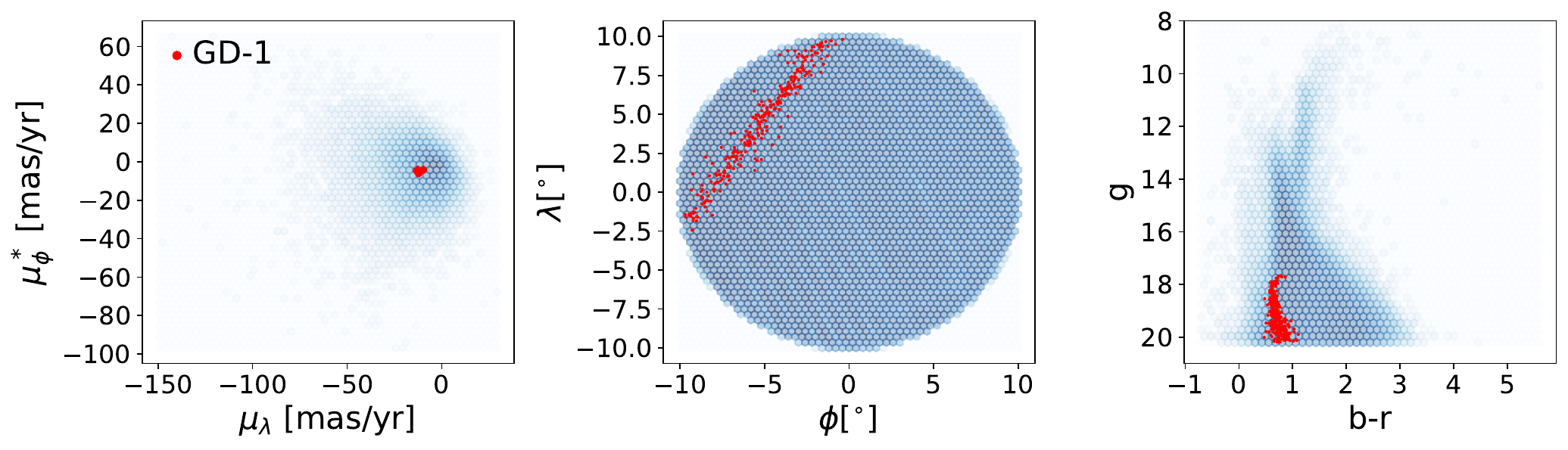}
\caption{Stars catalogued as members of the stellar stream GD-1 \citep{Price_Whelan_2018} is marked in red against the blue background stars in the patch with coordinates $(\alpha,\delta)=(148.6^\circ,24.2^\circ)$. Left: proper motion space; Center: angular space on the sky; Right: color-magnitude space.}
\label{fig:GD1}
\end{figure*}

\begin{figure}[!ht]
\centering
\includegraphics[width=0.75\linewidth]{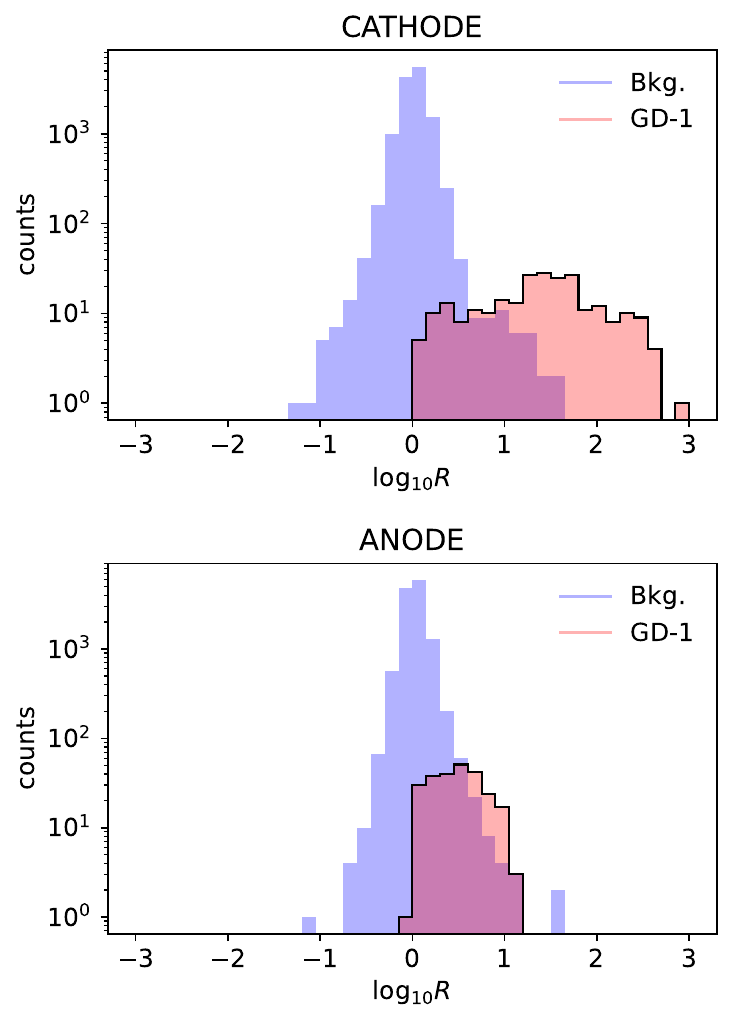}
\caption{The log $R$ values, as calculated by \cathode{} and \anode{}, of the signal stars and background stars in the signal region defined by $\mu_\lambda = [-17,-11]$ in the patch with coordinates $(\alpha,\delta)=(148.6^\circ,24.2^\circ)$, containing the GD-1 stream.}
\label{Cathode_vs_Anode_GD1_logR}
\end{figure}

\begin{figure}[!ht]
\centering
\includegraphics[width=\linewidth]{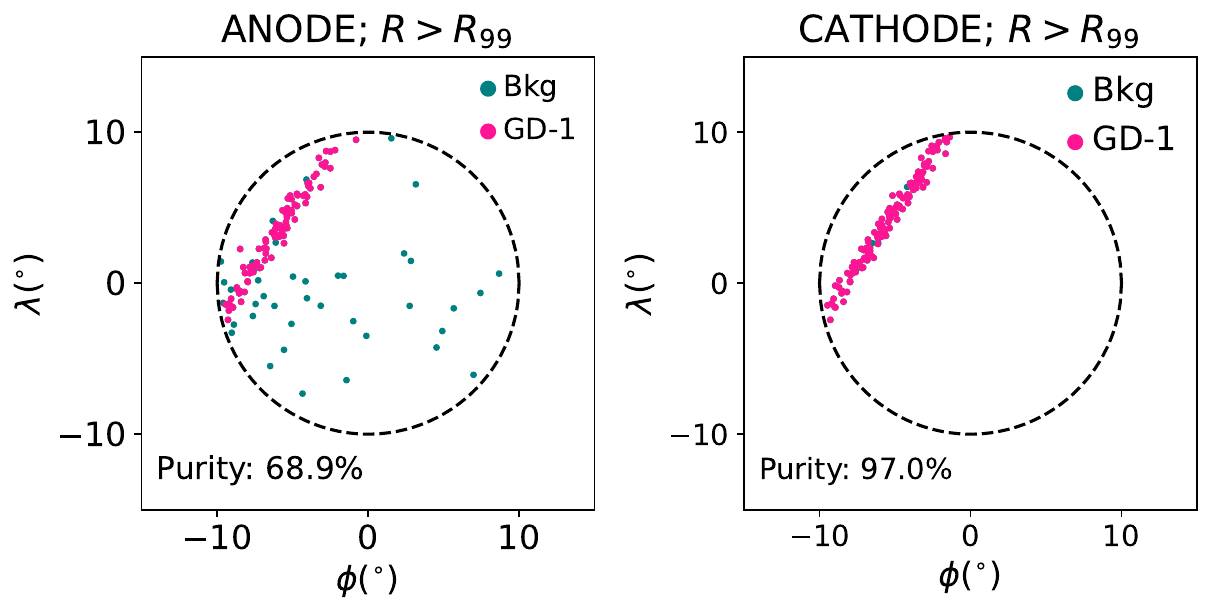}
\caption{The 1\% most anomalous stars, according to \cathode{} and \anode{}, in in the signal region defined by $\mu_\lambda = [-17,-11]$ in the patch with coordinates $(\alpha,\delta)=(148.6^\circ,24.2^\circ)$, containing the GD-1 stream. The color shows whether the stars are actually part of GD-1 (magenta) or if they are background stars (teal). The purity (as given in the lower left corner of the plots) denotes how many of the chosen stars are actually signal stars.}
\label{Cathode_vs_Anode_GD1_position}
\end{figure}

\subsection{Initial anomaly detection tests on GD-1}
\label{initial_GD1_tests}

We first compare \cathode{} and \anode{} as the basis for our stream detection algorithm using the well known and very bright stream GD-1 \citep{Grillmair_2006}, first identified in 2006. The stream is believed to have originated from a globular cluster that was tidally stripped by the Milky Way \citep{Koposov_2010} . As the stream has been further studied and mapped \citep{Price_Whelan_2018}, it has been discovered that the stream has substructure. There are two gaps, a ``spur'', and a blob outside of the main stream, some of which might be the result of an encounter with a dark matter substructure \citep{Bonaca_2019}. Figure~\ref{fig:GD1} shows how this stream is localized in the six dimensions $(\phi,\lambda,\mu_{\phi}^*,\mu_{\lambda},g,b-r)$ for one patch in the \textit{Gaia} sky.

For the evaluation, we chose a signal region where the stream stars are very distinct: in the patch centered on $(\alpha,\delta)=(148.6^\circ,24.2^\circ)$, we use the signal region $\mu_\lambda = [-17,-11]$. The probability output from the \cathode{} classifier step, $p$, is converted to an $R$ value using 
\begin{equation}
R = \frac{p}{1-p}.
\end{equation}
This allows comparison with the likelihood ratio provided by \anode{}, Eq.~\eqref{eq:Rofx}. We use the stream catalog from \cite{Price_Whelan_2018} to assign stars as members of GD-1.

In Figure~\ref{Cathode_vs_Anode_GD1_logR}, we plot the $R$ values assigned to stars that are ``signal'' (i.e., members of GD-1) and background using both \cathode{} and \anode{}. As can be seen, \cathode{} achieves a greater separation between the bulk of the signal and background stars. Taking the 1\% most anomalous stars for concreteness, we can achieve a signal-rich sample with a purity of 97\% using \cathode{} but only 69\% using \anode{}. This can be seen quite clearly by eye in Figure~\ref{Cathode_vs_Anode_GD1_position}, which shows the stars that survive the optimal \anode{} and \cathode{} selections. This demonstration shows that, at least in the case of high signal-to-background ratio streams such as GD-1, \cathode{} is more sensitive when it comes to detecting signal stars and reducing the false positive rate.

\section{The \vm{} method}
\label{sec:method}

\subsection{A brief review of Via Machinae}

Here we give a brief review of the full \vm{} algorithm for model-agnostic stream-finding. For more details we refer the reader to~\citep{Shih:2023jfv}. 

\begin{enumerate}

    \item Stellar streams are kinematically cold, which means that the stars within a stream are concentrated in proper motion. Therefore, the first step is to define signal regions (SRs) in each patch as proper motion intervals of 6~mas/yr,  spaced 1~mas/yr apart:
    \begin{equation}
    [\mu_{\lambda, {\rm min}},\mu_{\lambda,{\rm max}}] = \dots, [-5,1],[-4,2],[-3,3],[-2,4], \dots
    \end{equation}
    The width of the SRs is motivated by the typical proper motion extent of stellar streams previously discovered in {\it Gaia} data by \textsc{Streamfinder}.
    Only SRs with more than 20,000 and less than 1,000,000 stars are considered. For each SR (proper motion window), the full complement (all the stars in the patch not in the proper motion window) serves as the SB region.
    Note that there are two orthogonal proper motions --- latitudinal ($\mu_\lambda$) and longitudinal ($\mu_\phi^*$) --- in each patch; we separately consider SRs defined in terms of each.
    
    \item In each SR, a resonant anomaly detection method (either \anode{} or \cathode{} in this work) is trained to learn anomaly scores $R(x)$ for the stars in each SR. $R(x)$ is a measure of how overdense a region of phase space is relative to the data-driven background estimate. 
    
    \item Each SR is further divided into regions of interest (ROIs), defined using the orthogonal proper motion coordinate as illustrated in Figure~\ref{fig:regions}.  For each signal region, each ROI occupies ranges of the orthogonal proper motion coordinate ($\phi$ for $\lambda$ and vice versa) with 6~mas/yr:
    \begin{equation}
    [\mu_{\perp, \rm min},\mu_{\perp, \rm max}] = \dots, [-5,1],[-4,2],[-3,3],[-2,4], \dots
    \end{equation}
    ROIs
    containing globular clusters and dwarf galaxies are removed. In the remaining ROIs a fiducial region is defined by selecting stars that satisfy the following requirements:
    \begin{equation}
    g <20.2,\quad
    |\rho| <10^\circ,\quad
    b-r \in [0.5,1],\quad
    |z_{95}| >2~\mathrm{kpc},
    \end{equation}
    where $\rho$ is the angular distance from the center of the patch and $|z_{95}|$ is the Galactocentric altitude (distance to the center of the Galactic disk). Like the width of the SRs and ROIs, these requirements are motivated by the typical structure of known streams.
    In each ROI, we select the 100 stars with the highest $R$ values obtained from the resonant anomaly detection.\footnote{ROIs with less than 200 stars after the fiducial cut are discarded.} 
    
    \item Although the 100 highest $R$ stars in each ROI may be anomalous relative to the background, they might not be stream-like. To identify ROIs containing stream-like structures, the Hough transform \citep{Hough,Duda_Hart} is used to automate the process of finding stars that lie on a line in position space which passes at an angle $\hat\theta_{\rm line}$ through the patch and makes a closest approach to the center of angular distance $\hat\rho_{\rm line}$. The line-finding also returns a measure of the statistical significance of the best-fit line, $\sigma_{\rm line}$. 
    
    \item The next step is to combine the lines in overlapping, independent ROIs into protoclusters, to enhance their significance. Since each SR corresponds to an independent run of the anomaly detector, ROIs within a single SR are not independent, but ROIs from different SRs are. Once the ROIs are combined, a new line significance will be calculated by running the line finder step on the concatenation of the stars in the constituent ROIs. If this new significance is higher than the line significances from the individual ROIs, and if they fulfill a proper motion distance threshold,\footnote{This threshold was tuned based on inspections of known streams, false positives in \textit{Galaxia} and high-significance \textit{Gaia} stream candidates.} then the ROIs and their lines are combined into a new protocluster. Protoclusters that overlap with additional ROIs and previously-combined protoclusters are themselves combined, using the same algorithm. The protoclustering process continues until there are no more ROIs in the patch that can be clustered together. 
    
    \item Since non-independent ROIs cannot be clustered together, the protoclustering process can result in several protoclusters that are duplicates of one another. Two protoclusters are considered duplicates if more than $40\%$ of the line stars in one of the clusters are also included in the other cluster's line stars. A set of duplicate protoclusters are referred to as a \textit{protostream}. The significance of the protostream is the highest significance of any of its protoclusters. 
    
    \item Finally, protostreams are merged over patches, resulting in \textit{stream candidates}. This is the final output of the stream-finding algorithm. In order to be merged into a stream candidate, two protostreams must locally align both in proper motion and line direction. This condition is tested for each protocluster pair $\mathcal{C}^{(1)}_i, \mathcal{C}^{(2)}_j$ between two protostreams $\mathcal{P}^{(1)}, \mathcal{P}^{(2)}$ in overlapping patches. The final significance $\sigma(\mathcal{S})$ of the stream is the quadrature sum of the significances $\sigma(\mathcal{P})$ of the component protostreams.

\end{enumerate}

\subsection{Re-optimizing Via Machinae}
\label{sec:retuning}

In this technical subsection, we describe a re-optimization of the \vm{} algorithm, some of which is specific to \vmc, and some of which applies to both \vmc{} and \vma{}. This section will not be self-contained but will assume prior knowledge of VM2. The reader that is interested primarily in the new \vmc{} and \vma{} stream candidates is encouraged to skip to Section~\ref{sec:results}.

\subsubsection{Protostream matching}
\label{sec:protostream_matching}

A key criterion when re-tuning the \vm{} cuts for \cathode{} will be to preserve the performance of the algorithm on the top six known streams found with VM2 (GD-1; Gaia-1 \citep{2018MNRAS.481.3442M}; Jhelum \citep{2018ApJ...862..114S}; Leiptr, Fjorm and Ylgr \citep{2019ApJ...872..152I}). For this reason, we will find it helpful to match the protostreams from \vmc{} to their counterparts in \vma{} corresponding to the known streams.

\begin{figure}[h]
    \centering
    \includegraphics[width=0.65\linewidth]{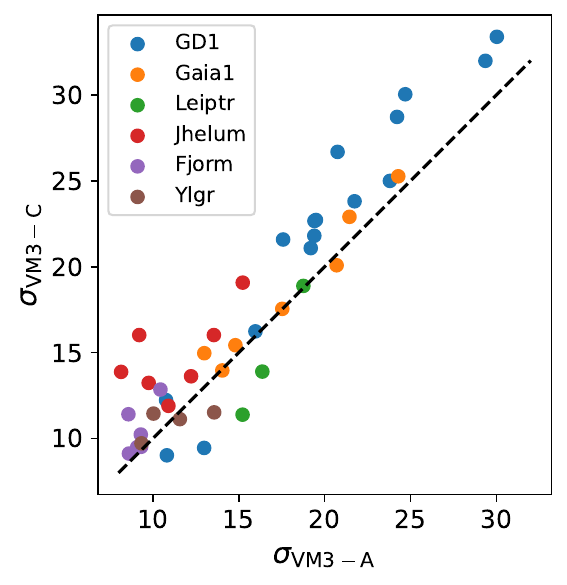}
    \caption{\vma{} vs \vmc{} significances of the 42 protostreams corresponding to the top six known streams found by \vma{}.}
    \label{fig:ps_sig_matched}
\end{figure}

Given the high significance of the protostreams from the top six known streams found with \vma{},  it is straightforward to match them to protostreams found with \vmc{}. The matching algorithm is as follows: For each \vma{} protostream coming from a known stream, select its highest significance protocluster (defined as ${\mathcal C}_{\rm A}$).  Loop through all the \vmc{} protostreams in the same patch. For each protostream ${\mathcal P}_{\rm C}$, loop through all of its protoclusters ${\mathcal C}_{\rm C}$. Count the fraction of unique stars in ${\mathcal C}_{\rm A}$ that are found in ${\mathcal C}_{\rm C}$ and maximize this fraction over all the protoclusters in ${\mathcal P}_{\rm C}$. If this fraction is greater than 0.25, call it a match:
\begin{equation}
f_{\rm overlap} \equiv \max_{{\mathcal C}_{\rm C} \in
{\mathcal P}_{\rm C}}
\frac{{|{\mathcal C}_{\rm A}\cap {\mathcal C}_{\rm C}|}}{{ |{\mathcal C}_{\rm A}|}}>0.25
\end{equation}
This cut was tuned by eye and found to give good results, in that each \vma{} protostream has exactly one matching \vmc{} protostream that is a clear by-eye match in the feature space. 

Altogether there are 42 protostreams from the known streams found by \vma{}, and all of these are confirmed by \vmc{}.
A scatter plot of the \vmc{} vs. \vma{} significances is shown in Figure~\ref{fig:ps_sig_matched}. 
We see that there is an approximately linear correlation between the \vmc{} and \vma{} significances; performing a simple linear regression fit yields a best-fit slope and intercept of $1.08\pm0.06$ and $0.31\pm1.03$ respectively. 
Evidently, \vmc{} increases the significances of the known-stream protostreams relative to \vma{} by $\sim\!10\%$, but there is some spread in the outcome. Interestingly, the correlation seems to become tighter as the significance increases. This is to be expected from optimal-anomaly-score-based methods: when there is enough signal (in this case stream stars) present, these methods are very powerful and certain to enhance the signal significance. But when the initial signal level is too low, these methods lose signal sensitivity and can be subject to statistical fluctuations.

\subsubsection{Increasing the protocluster significance cut}
\label{sec:protocluster_cut_increase}

In VM2 the protocluster significance was required to be $\sigma({\mathcal P})>8$, as those protoclusters with lower significance were all consistent with being false positives. This was based on a comparison of the cumulative distribution of protocluster significance between {\it Gaia} and {\it Galaxia}. 

\begin{figure}[h]
    \centering
    \includegraphics[width=0.85\linewidth]{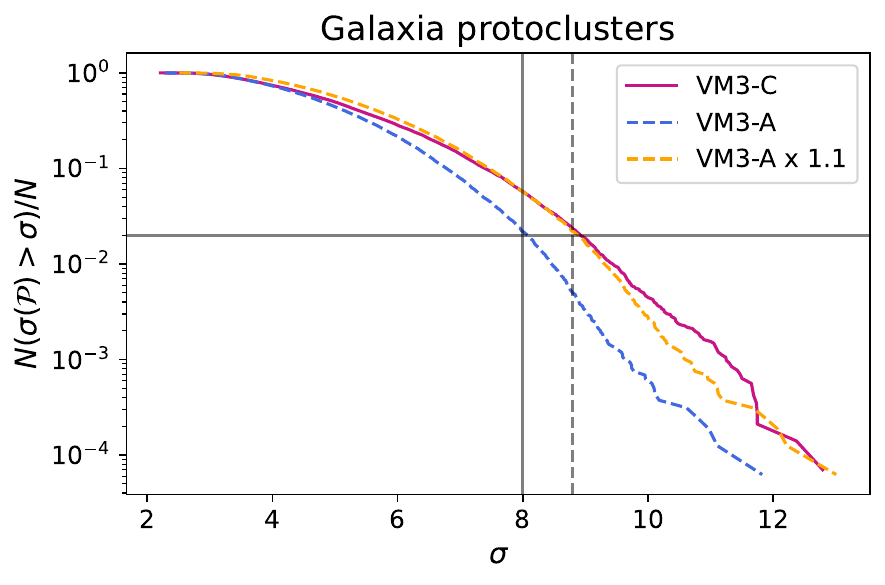}
    \caption{The survival function of Galaxia protoclusters vs their significance for \vma{} and \vmc{}. The cutoff at $\sigma=8$ corresponds to an \vma{} survival fraction of 2\%; the same survival fraction for \vmc{} corresponds to a significance of $\sigma=8.8$. The two survival functions match well when the \vma{} significances are rescaled by 10\%.}
    \label{fig:survival_comp}
\end{figure}

For \vmc{} we find it beneficial to increase this cut to $\sigma({\mathcal P})>8.8$. This is motivated by the comparison of the survival functions with respect to the protocluster significance $s(\sigma) = N(\sigma({\mathcal P})>\sigma)/N$
between \vma{} and \vmc{}. This is shown in  Fig.~\ref{fig:survival_comp}. We see that the \vma{} survival function matches the \vmc{} one well when the protocluster significances for \vma{} are rescaled by 10\%. The survival probability at $\sigma({\mathcal P})=8$ is 2\% for \vma{}; the corresponding $\sigma({\mathcal P})$ for the same survival probability for \vmc{} is indeed $\sigma({\mathcal P})=8.8$. 
So this motivates us to increase the cut on protocluster significances by the same factor, from 8 to 8.8.

\subsubsection{Removing edge protostreams}

By visual inspection of high-significance {\it Galaxia} protostreams, we noticed that a disproportionate number of them are located at the edges of patches for \vmc{}, and to a lesser extent for \vma{}. Thus we are motivated to remove edge protostreams from our sample, in order to reduce the false positive rate. According to the {\it Galaxia} sample, it makes sense to define an edge protostream as those protostreams whose best-fit line parameters have radius $\hat\rho>7.8^\circ$.

\begin{figure}[h]
    \centering
    \includegraphics[width=0.75\linewidth]{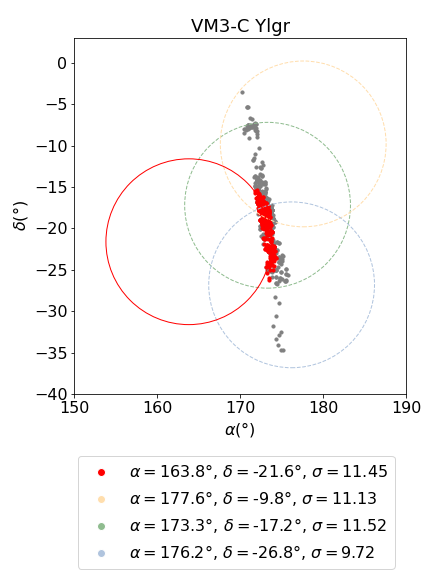}
    \caption{Example of a \textit{Gaia} edge protostream from the known stream Ylgr/Gaia-3. The edge protostream is marked in red, while the other protostreams in the stream are marked in grey. The respective patches in which the protostreams were found are illustrated with circles. The locations of the patches and the significances of the protostreams are indicated in the legend.}
    \label{fig:ps_edge_ylgr}
\end{figure}

While many of the edge protostreams might be false positives, not all of them are: in Figure~\ref{fig:ps_edge_ylgr} we show an example of a 
\vmc{} protostream coming from the known stream Ylgr/Gaia-3 which is high significance ($\sigma=11.5$) and lies at the edge of its patch ($\hat\rho_{\rm line}=8.4^\circ$). 

Our prescription for dealing with edge protostreams is as follows. If an edge protostream is merged with any other non-edge protostream in the final stream clustering step, we keep it (this criteria is easily satisfied in the Ylgr/Gaia-3 example). If an edge protostream occurs by itself in the final list of stream candidates, we consider it a likely false positive and remove it.

\subsubsection{Re-tuning the $f_{\rm dim}$ cut}

In VM2, by inspecting the highest significance \textit{Galaxia} (proto)streams, it was discovered that they were all predominantly composed of bright stars, whereas the \textit{Gaia} protostreams tended to have a higher fraction of dim stars. Thus to further reduce the false positive rate, a cut on the fraction of dim stars $f_{\rm dim}$ was found to be helpful. Here $f_{\rm dim}$ is defined as the fraction of stars in a protostream with magnitude $g>g_c$, where $g_c$ is a tunable hyperparameter.

We find that it is necessary to re-tune the cut to optimize for \cathode{}. In doing so, we wish to satisfy two constraints: (1) we want to preserve the performance of \vm{} on the six highest-significance known-streams from \vma: GD-1, Gaia-1, Leiptr, Jhelum, Fjorm, Ylgr; and (2) we want to minimize the significance of the highest-significance \textit{Galaxia} protostream. 

\begin{figure}[h]
    \centering
    \includegraphics[width=0.95\linewidth]{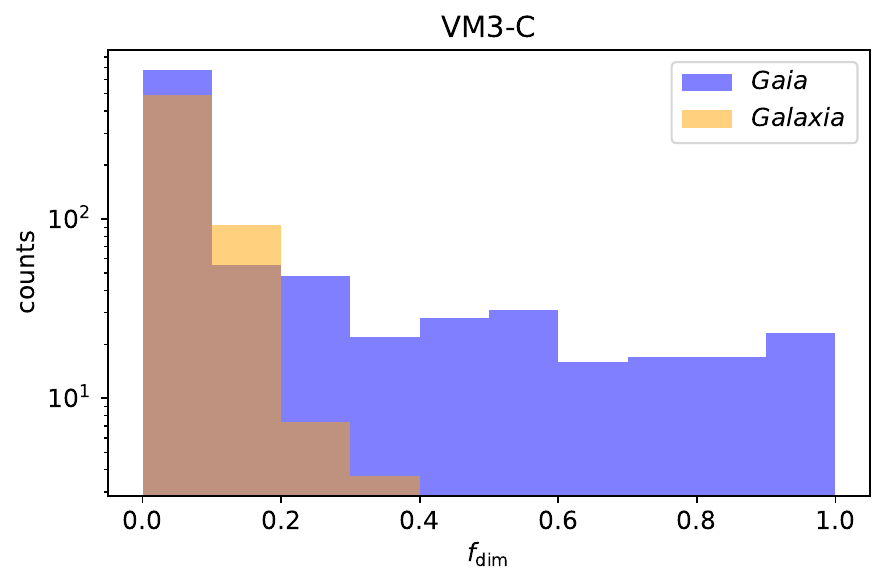}
    \caption{Distribution of $f_{\rm dim}$ with $g_c=19.1$ for {\it Gaia} and {\it Galaxia} protostreams in \vmc{}. }
    \label{fig:fdim_CATHODE}
\end{figure}

We tune the $f_{\rm dim}$ cut to minimize the significance of the highest significance {\it Galaxia} protostream while keeping all the protostreams corresponding to the top 6 known {\it Gaia} streams. 
This tuning reproduces $g_{c}=18.4$ and $f_{\rm dim}=0.5$ for \vma, with the significance of the highest significance \textit{Galaxia} protostream $\sigma=8.86$. For \vmc{} it produces $g_{c}=19.1$, $f_{\rm dim}=0.25$ and a maximum \textit{Galaxia} significance of $\sigma=8.92$. Figure \ref{fig:fdim_CATHODE} shows the distribution of $f_{\rm dim}$ with $g_{c}=19.1$ for  {\textit Gaia} and {\it Galaxia} protostreams found by \vmc. As in \vma{}, we see that the $f_{\rm dim}$ distribution of {\textit Gaia} is well-described by  {\it Galaxia}  at lower $f_{\rm dim}$, suggesting that these are all predominantly false positives. 

\subsubsection{Loosening the automated stream clustering criteria}
\label{sec:streamclustering}

The nominal stream clustering criteria from \cite{Shih:2023jfv} were as follows: for each pair of protostreams ${\mathcal P}_A$ and ${\mathcal P}_B$ in adjacent patches, consider all pairs of protoclusters ${\mathcal C}_A\in {\mathcal P}_A$ and ${\mathcal C}_B\in {\mathcal P}_B$. If any pair of protoclusters $({\mathcal C}_A,\,\,{\mathcal C}_B)$ has more than $3^\circ$ of overlap, calculate the fraction of one protocluster's line stars that are also present among the other protocluster's line stars, $f_{\rm line}$. If $f_{\rm line}>0.4$, compute the difference between their best-fit Hough line parameters, $\Delta\theta_{\rm line}$ and $\Delta\rho_{\rm line}$.
Require that these differences, averaged over these pairs of protoclusters, satisfy 
$|\langle\Delta\theta_{\rm line}\rangle|<9^\circ$ and $|\langle\Delta\rho_{\rm line}\rangle|<1^\circ$. 

We noticed through a by-eye inspection that several \vmc{} and \vma{} stream candidates could apparently be merged further. 
So for this work we have opted to loosen the automated stream clustering requirements as follows. Either the protostreams satisfy the above criteria, \textit{or} they satisfy the following looser criteria: for all pairs of protoclusters with more than  $3^\circ$ of overlap and $f_{\rm line}>0.1$, require $|\langle\Delta\theta_{\rm line}\rangle|<14^\circ$ and $|\langle\Delta\rho_{\rm line}\rangle|<1^\circ$. The \vma{} results diverge from that of VM2 due only this choice and the removal of edge protostreams.

\section{Results}
\label{sec:results}

\subsection{Estimating the expected number of false positives}

Clustering together the protostreams that pass the re-tuned $f_{\rm dim}$ cut and removing singleton edge protostreams results in \vmc{} identifying 103
\textit{Gaia} stream candidates and just 1 in \textit{Galaxia}. This is to be compared with 202 \textit{Gaia}
vs.~12 \textit{Galaxia} stream candidates
with \vma{}. 

\begin{figure}[h]
    \centering
    \includegraphics[width=\linewidth]{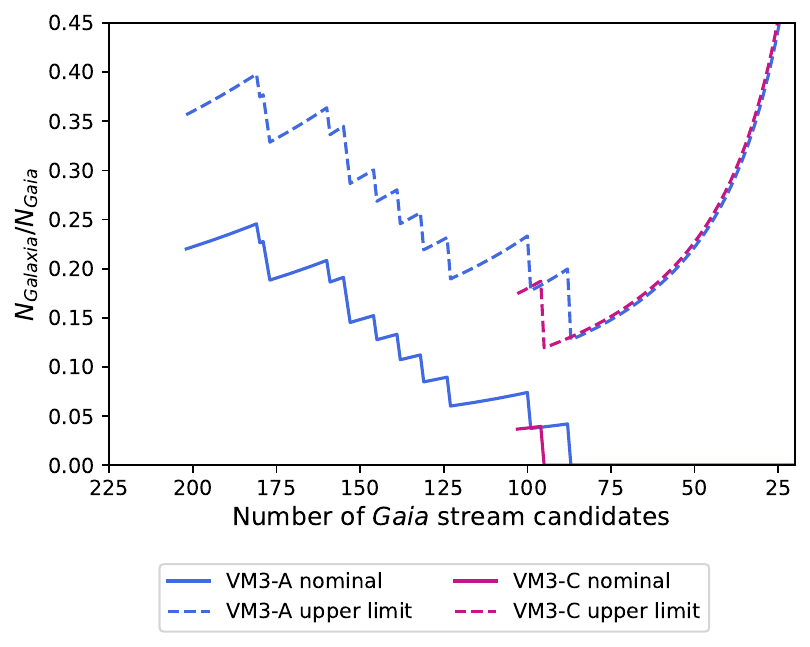}
    \caption{$N_{Galaxia}(\sigma({\mathcal S})>\sigma_c)/N_{Gaia}(\sigma({\mathcal S})>\sigma_c)$ vs.\ $N_{Gaia}(\sigma({\mathcal S})>\sigma_c)$ for \vma{} and \vmc{}, as the cut on stream significance $\sigma_c$ is varied.}
    \label{fig:fpr_vs_sig}
\end{figure}

As in VM2, we use a cut on stream significance to purify the stream candidates and minimize the false positive rate. Figure~\ref{fig:fpr_vs_sig} shows $N_{Galaxia}(\sigma({\mathcal S})>\sigma_c)/N_{Gaia}(\sigma({\mathcal S})>\sigma_c)$ vs.\ $N_{Gaia}(\sigma({\mathcal S})>\sigma_c)$ as the cut $\sigma_c$ on stream significance is varied. Here $N_{Galaxia}$ has been rescaled by 163/44 to account for the smaller number of patches used for the {\it Galaxia} scan. After a cut that removes all the \textit{Galaxia} streams, there are 95 remaining stream candidates in \vmc{} and 87 in \vma{}. Using Poisson statistics, the upper limit on how many false positives can be expected in the \textit{Gaia} data after this cut is  12.0\% for \vmc{} and 12.8\% for \vma{}.
In summary, after taking the Poisson statistics into account we expect that (at least) 84 of the \vmc{} stream candidates are real streams, compared to 76 in \vma{}. This represents an increase by over 10\% in the number of high-confidence stream candidates.

\subsection{Stream candidates}
\label{sec:stream_candidates}

We next consider the specific stream candidates identified by \vmc{} and \vma{}. Because of the large number of stream candidates found, here we restrict ourselves to discussing streams that fulfill any of the following criteria:\footnote{The complete set of stream candidates will be made public after the manuscript is accepted for publication.}
\begin{enumerate}
    \item correspond to a known stream in the \texttt{galstreams} library,
    \item are found in both \vmc{} and \vma{},
    \item contain at least two protostreams,
    \item contain a single protostream but have a higher significance than a multi-protostream stream.
\end{enumerate}

In order to evaluate the second criterion, we employ a mixture of the matching procedure from Section~\ref{sec:protostream_matching} (looking for protostreams within the \vma{} and \vmc{} stream candidates that have a high degree of overlap) and simple by-eye inspection of the stream candidates in position, proper motion, and photometry space.

\subsubsection{Performance on known streams}

\begin{table*}
\begin{center}
\renewcommand\arraystretch{1.3}
\begin{tabular}{|>{\centering\arraybackslash}m{0.18\linewidth}|>{\centering\arraybackslash}m{0.08\linewidth}|>{\centering\arraybackslash}m{0.08\linewidth}|>{\centering\arraybackslash}m{0.07\linewidth}|>{\centering\arraybackslash}m{0.07\linewidth}|}
\hline 
\textbf{Known stream} &
\textbf{\vmc{}} & \textbf{\vma{}} & 
\textbf{$\sigma_{\rm VM3-C}$} 
 & \textbf{$\sigma_{\rm VM3-A}$}   \\
\hline 
\hline 
GD-1 & 1 & 1 & 93.65 &	83.43      \\
\hline 
Gaia-1 & 2 & 2 &  50.10	& 48.69      \\
\hline 
Jhelum & 3 & 4 & 36.27 & 29.07     \\
\hline 
Fjorm & 4 & 5  &   31.12 &  22.63    \\
\hline 
Sagittarius & 7, 19, 51 & 9, 17, 41, 74, 126, 153, 173, 177	& 30.67 & 29.97  \\
\hline 
Gaia-8 & 8, 55 & 10, 11, 16 	& 26.77  & 26.98  \\
\hline 
Leiptr  & 6 & 3  & 26.07	& 29.18     \\
\hline 
Ylgr  & 9 & 6 &  21.96	& 22.49   \\
\hline
 ACS & 22 & 72 &  14.03	& 9.06   \\
\hline 
Lethe & 39 & 98 & 10.69 & 8.72  \\
\hline 
Tri-Pis B12  & 44 & - & 10.15 & - \\
\hline
M68/Fjorm & 87 & - &  9.10  & -  \\
\hline
Cetus-Palca-T21 & 88 & - &   9.07 &  -  \\
\hline
Gaia-9 & - & 45 & -  & 9.57   \\
\hline
Gaia-6  & - & 85 & -  & 8.89 \\
\hline
 possible continuation of Gaia-6 & - & 164 & -  & 8.19  \\
\hline
\end{tabular}
\caption{Previously known streams found with \vma{} and/or \vmc{}. The numbering in the first two columns refers to the order of the stream candidates in terms of significance. The reported significance values for Sagittarius and Gaia-8 are calculated by adding the significances of the individual candidates in quadrature.}
\label{tab:stream_comparison_known}
\end{center}
\end{table*}

The previously-known streams found by \vmc{} and \vma{} are
presented in Table~\ref{tab:stream_comparison_known}. The ones found in both methods are: GD-1
\citep{Grillmair_2006}; Gaia-1 and Gaia-3
\citep{2018MNRAS.481.3442M}; Jhelum \citep{2018ApJ...862..114S}; Leiptr,
Fjorm and Ylgr (which appears to be the same object as Gaia-3)
\citep{2019ApJ...872..152I}; Sagittarius \citep{2002ApJ...569..245N,2003ApJ...599.1082M}; Gaia-8 \citep{2021ApJ...914..123I, ibata2023chartinggalacticaccelerationfield}; ACS~\citep{Grillmair_2006} and Lethe~\citep{Grillmair_2009}. 

Furthermore, \vmc{} (but not \vma{}) found  Cetus-Palca \citep{Yanny_2009,Newberg_2009,Shipp_2018,Thomas_2021}, and what appears to be the globular cluster M68 (the progenitor of Fjorm \citep{10.1093/mnras/stz1790}). 
\vmc{} also found what could possibly be Triangulum-Pisces \citep{Bonaca_2012,Martin_2013}, although this stream candidate shows some characteristics of a globular cluster. 

Finally, \vma{} (but not \vmc{}) found both Gaia-6 \citep{2019ApJ...872..152I} and Gaia-9 \citep{2021ApJ...914..123I}. \vma{} also found a possible continuation of Gaia-6. Although this stream candidate has a lower significance, its agreement with the main part of Gaia-6 in both position space and proper motion space is striking enough that we believe that there is a high probability that this is in fact part of Gaia-6. We have not found this continuation of Gaia-6 in the literature, which means that this could be the first discovery of this extension.

All these streams are shown in Figures~\ref{fig:known_streams1}, \ref{fig:known_streams2}, \ref{fig:known_streams3}, \ref{fig:known_streams4} and \ref{fig:known_streams5} in Appendix~\ref{sec:knownstreams}. In all plots, stream tracks from the \texttt{galstreams} Milky Way stellar stream library \citep{Mateu_2023} have been overlaid. Note that these stream tracks show the path of the stream on the sky, not the individual stars nor the thickness of the stream. This makes Sagittarius, a very thick stream, look like it is displaced from the detected fragments in Figure~\ref{fig:known_streams2}. In reality, these fragments are well within the extent of the stream.

An interesting find among the lower significance streams of \vmc{} is Lethe~\citep{Grillmair_2009}, shown in Figure~\ref{fig:known_streams3}. The \texttt{galstreams} library lacks proper motion measurements for Lethe, which is why this is not shown in the middle panel of Figure~\ref{fig:known_streams3}. 
To our knowledge, this could be the first measurement of the proper motions of Lethe member stars. This shows how re-discovery of known streams using the \vm{} method can aid in expanding the knowledge of the stellar population properties of such streams.

As can be seen from Figure~\ref{fig:known_streams1}, \vmc{} also confirms the \vma{} extensions of Gaia-1  and Jhelum compared to \textsc{Streamfinder}, first noted in \cite{Shih:2023jfv}.

\subsubsection{Discarded candidates}
Of the 79 stream candidates in \vmc{} with with $\sigma>8.92$  (the significance cut required to remove all \textit{Galaxia} stream candidates) that do not correspond to any known stream in the \texttt{galstreams} library, five (candidates 23, 26, 36, 37 and 40) are discarded since the stars do not form a proper line. This is a failure mode of the line-finding algorithm, which can happen when anomalous stars in two different regions of a patch result in a high-significance line, despite there being no line of stars between these two groups. For example, stream candidate 23 contains 55 stars, 51 of which are located in the region $\ell~=~[68.26^\circ, 68.41^\circ], b~=~[34.79^\circ, 34.92^\circ]$, which agrees very well with the reported location $(\ell,b)=(68.34^\circ,34.89^\circ)$ of the globular cluster M92 in \cite{10.1093/mnras/stab1475}. Note however that the proper motions of these stars are significantly off from the reported value, which means that we will not claim that this stream candidate is actually M92. This is illustrated in Figure~\ref{fig:no_stream} in Appendix~\ref{sec:knownstreams}. The fact that this passed through as a stream candidate, despite the method from \cite{Shih:2023jfv} of filtering out globular clusters being employed, illustrates the importance of carefully examining the resulting candidates from this type of stream-finding method. 

Of the 65 stream candidates in \vma{} with with $\sigma>8.86$  (the significance cut required to remove all \textit{Galaxia} stream candidates) that do not correspond to any known stream in the \texttt{galstreams} library, no candidates had to be discarded.

\subsubsection{New streams}

\begin{table*}
\begin{center}
\renewcommand\arraystretch{1.3}
\begin{tabular}{|>{\centering\arraybackslash}m{0.1\linewidth}|>{\centering\arraybackslash}m{0.1\linewidth}|>{\centering\arraybackslash}m{0.1\linewidth}|>{\centering\arraybackslash}m{0.1\linewidth}|>{\centering\arraybackslash}m{0.1\linewidth}|}
\hline 
\textbf{VM3} & \textbf{\vmc{}} & \textbf{\vma{}} & 
\textbf{$\sigma_{\rm VM3-C}$} & 
 \textbf{$\sigma_{\rm VM3-A}$}  \\
\hline 
1 & 5 & 28 & 28.95 & 10.50 \\ 
\hline 
2 & 14 & 13, 47  & 17.84	 & 16.83 \\
\hline 
3 & 16 & 8  &  15.96	& 17.34 \\ 
\hline 
4 & 21 & 19  &  14.15	& 12.52  \\ 
\hline 
5 & 24 & 22  &  13.47 & 12.41 \\ 
\hline 
6 & 27 & 53, 144  & 13.05 & 12.58 \\ 
\hline 
7 & 29 & 40 & 12.80 & 9.76  \\ 
\hline 
8 & 38 & 136  & 10.80 & 8.34 \\ 
\hline 
9 & 48 & 43  & 9.94 & 9.68 \\ 
\hline 
10 & 54 & 30  & 9.65 & 10.44 \\ 
\hline 
11 & 56 & 163  & 9.59 & 8.20 \\ 
\hline 
12 & 57 & 24  & 9.59  &  11.89   \\
\hline 
13 & 65 & 158  & 9.43 & 8.22 \\ 
\hline 
14 & 67 & 148  & 9.38 & 8.27 \\ 
\hline 
15 & 93 & 21  & 8.95 & 12.44 \\ 
\hline 
16 & 11 & - &  20.55 & -  \\ 
\hline 
17 & 13 & - & 18.05 & -  \\ 
\hline 
18 & 15 & - & 16.87 & -  \\ 
\hline 
19 & 25 & - & 13.37   & - \\ 
\hline 
20 & 28, 83 &- & 15.73 & -  \\ 
\hline  
21& - & 14  &- &  13.72  \\ 
\hline 
22& - & 15  & - & 13.52    \\ 
\hline 
23& - & 18    & - & 12.72    \\ 
\hline 
24& - & 23    & - & 12.20    \\ 
\hline 
25& - & 25     & - & 11.60   \\ 
\hline 
\end{tabular}
\caption{A selection of new stream candidates not corresponding to any known stream in the \texttt{galstreams} library.
The numbering in the second and third columns refers to the order of the stream candidates in terms of significance. Since some stream candidates are believed to belong to the same object, we have gathered them together and then re-numbered our final result according to the column VM3. The significance values for clustered candidates are calculated by adding the significances of the individual candidates in quadrature.}
\label{tab:stream_comparison_new}
\end{center}
\end{table*}

After removing the known streams and the stream candidates that are failures of the line-finding algorithm as described above, we are left with 74 stream candidates from \vmc{} and 65 from \vma{}. Table~\ref{tab:stream_comparison_new} contains a selection of high-significance stream candidates without known stream matches, using the selection criteria described in Section~\ref{sec:stream_candidates} and excluding two larger clusters that will be discussed in the following section. Altogether, using this selection and a final by-eye merging of candidates that are deemed to be part of the same object, there are 25 potential new streams among \vmc{} and \vma{} combined. These can be divided into the following categories:
\begin{itemize}
    \item Found by both \vmc{} and \vma{}:\footnote{Here we allow lower-significance stream candidates from \vma{}, when it is a clear match to a \vmc{} candidate.} 15 streams, shown in ~\ref{fig:new_streams1}, \ref{fig:new_streams2}, \ref{fig:new_streams3} and \ref{fig:new_streams4} in Appendix~\ref{sec:newstreamcands}.
    \item Found only in \vmc{}: 5 streams, shown in Figures~\ref{fig:new_streams5} and \ref{fig:new_streams6}
    \item Found only in \vma{}: 5 streams, shown in Figures~\ref{fig:new_streams7} and \ref{fig:new_streams8}.
\end{itemize}

Beyond these candidates, and the candidates included in the two clusters we shall shortly turn to, there are 66 further candidates that are found by one method but not the other (35 in \vmc{} and 31 in \vma{}). This leads to a total collection of 93 new stream candidates. All of these are shown in a full-sky atlas in Figure~\ref{fig:streamatlas}, where the numbering of the high-confidence objects corresponds to the VM3 numbering in Table~\ref{tab:stream_comparison_new}. 

\subsubsection{Clusters of stream candidates}

\begin{table*}
\begin{center}
\renewcommand\arraystretch{1.3}
\begin{tabular}{|>{\centering\arraybackslash}m{0.2\linewidth}|>{\centering\arraybackslash}m{0.1\linewidth}|>{\centering\arraybackslash}m{0.1\linewidth}|>{\centering\arraybackslash}m{0.1\linewidth}|>{\centering\arraybackslash}m{0.1\linewidth}|}
\hline 
\textbf{VM3} & \textbf{\vmc{}} & \textbf{\vma{}} & 
\textbf{$\sigma_{\rm VM3-C}$} & 
\textbf{$\sigma_{\rm VM3-A}$}  \\
\hline 
 & 10  & 7 & 20.60 & 21.69 \\ 
\cline{2-5}
 & 18  & 56 & 14.85 & 9.38 \\ 
\cline{2-5}
 & 20 & - & 14.36  & - \\ 
\cline{2-5} 
 & 41 & - & 10.42 & - \\ 
\cline{2-5} 
 & 42  & - & 10.32 & - \\ 
\cline{2-5} 
 & 47 & - & 10.05 & - \\ 
\cline{2-5} 
 & 50  & 12 & 9.86 & 14.69 \\ 
\cline{2-5}
\multirow{2}{*}{\textit{Raritan}} & 53  & - & 9.68 & - \\ 
\cline{2-5}
 & 59 & 50 & 9.53 & 9.49 \\ 
\cline{2-5} 
 & 62  & 68 & 9.50 & 9.09 \\ 
\cline{2-5} 
 & 69 & - & 9.35  & - \\ 
\cline{2-5}
 & 70 & -  & 9.32 & - \\ 
\cline{2-5} 
 & 82 & 161  & 9.14 & 8.20 \\ 
\cline{2-5} 
 & 94 & - & 8.95 & - \\ 
\cline{2-5} 
 & - & 29 & - & 10.46 \\ 
\cline{2-5} 
 & - & 42 & - & 9.76 \\ 
\cline{2-5} 
 & - & 46 & - &  9.55 \\ 
\hline 
 & 12 & 20  & 19.60 & 12.45 \\ 
\cline{2-5} 
\multirow{2}{*}{\textit{Passaic}} & 17 & 77  & 14.94 & 8.99 \\ 
\cline{2-5} 
 & 32 & 62  & 11.98 & 9.23 \\ 
\cline{2-5} 
 & 91 & - & 8.96 & - \\ 
\hline 
\end{tabular}
\caption{The stream candidates that were clustered by eye into the two objects \textit{Raritan} and \textit{Passaic}. The combined significance (after adding the individual significances in quadrature) is 44.28 (35.25) for \textit{Raritan} in \vmc{} (\vma{}), and 28.83 (17.91) for \textit{Passaic} in \vmc{} (\vma{}).}
\label{tab:cluster_contents}
\end{center}
\end{table*}

\begin{figure*}
    \includegraphics[width=0.95\textwidth]{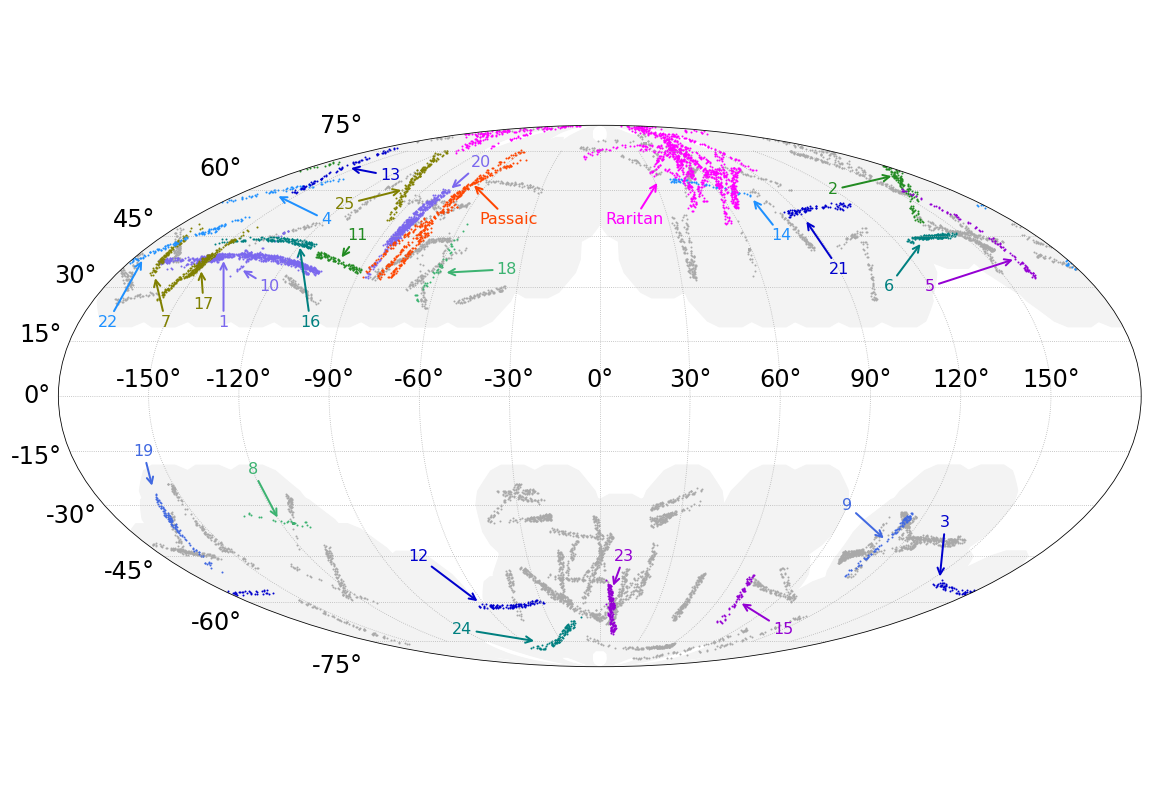}
    \caption{All-sky map in Galactic coordinates of the stream candidates found by \vmc{} with $\sigma>8.92$ and by \vma{} with $\sigma>8.86$ (the significance cut required to remove all \textit{Galaxia} stream candidates) that do not correspond to any known stream in the \texttt{galstreams} library, and excluding 5 stream candidates that were deemed to be failures of the line-finding algorithm. In total there are 74 stream candidates in \vmc{} and 65 in \vma{} fulfilling these requirements. After a final by eye-inspection leading to the grouping of several candidates into clusters, there are a total of 15 high-confidence stream candidates that are found by both \vmc{} and \vma{}, plus two larger clusters that are also found by both methods. A further 10 candidates were found by one method and but not the other. The two clusters are marked in magenta and orange, the other 25 candidates are drawn in various other colors and numbered according to the VM3 numbering in table~\ref{tab:stream_comparison_new}. The remaining 66 candidates are marked in grey and not numbered. The part of the sky that is covered by the patches used in this work is shaded in grey.}
    \label{fig:streamatlas}
\end{figure*}

\begin{figure*}
\centering
    \subfigure{\includegraphics[width=0.79\textwidth, valign=t]{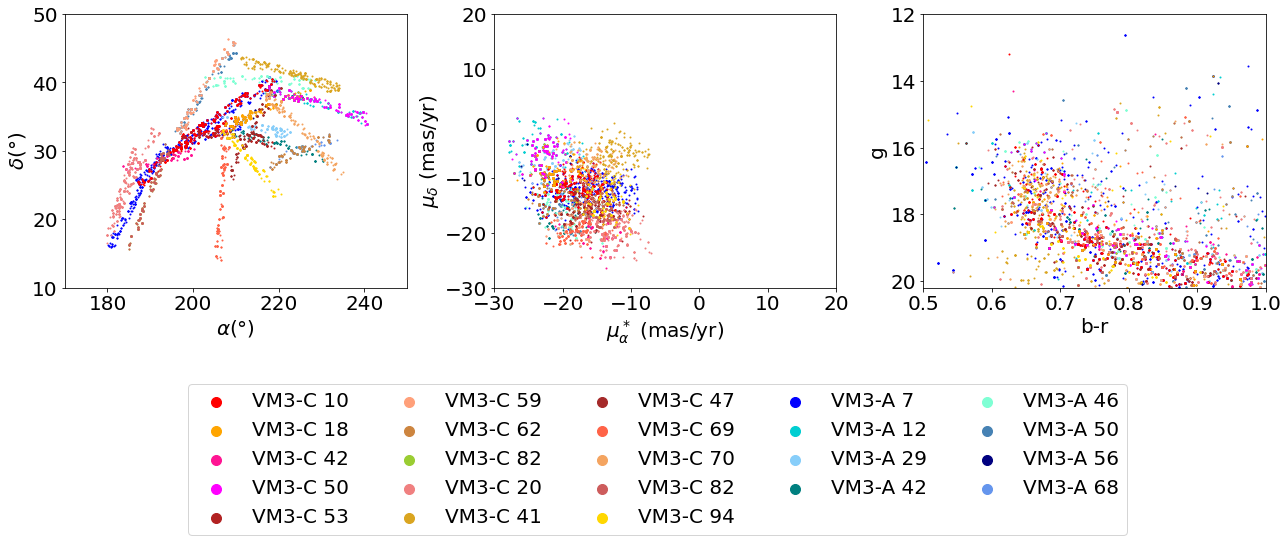}}    
    \subfigure{\includegraphics[width=0.79\textwidth, valign=t]{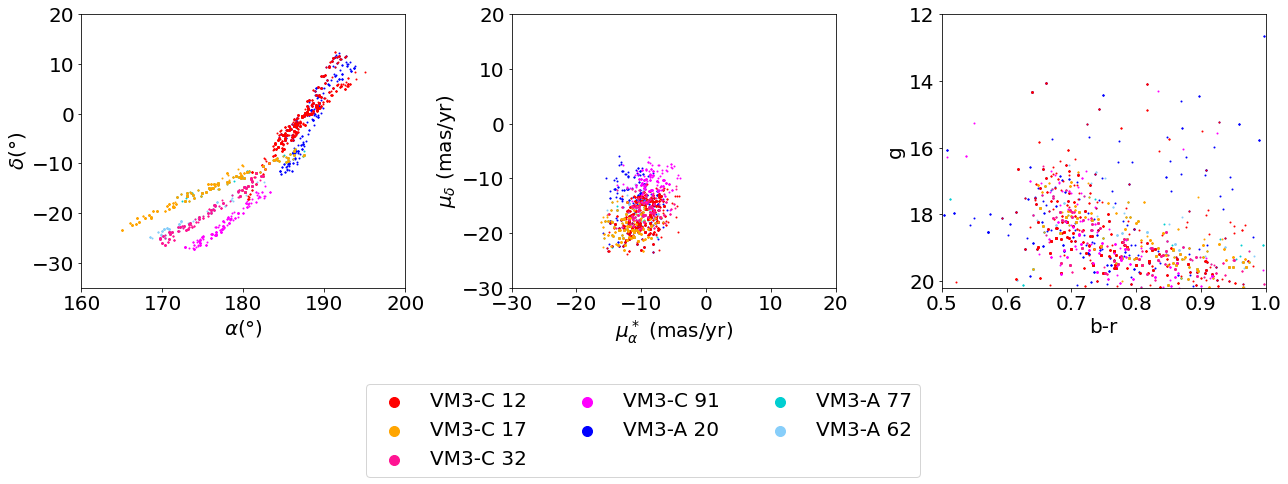}}
    \caption{A final by-eye clustering of stream candidates leads to the following two clusters: the \textit{Raritan} (top) consisting of 14 \vmc{} candidates and 9 \vma{} candidates, of which 6 overlap between the two methods, and the \textit{Passaic} (bottom) containing 3 candidates found by both \vmc{} and \vma{} and one further candidate found by \vmc{} only.}
    \label{fig:NorthClusters}
\end{figure*}

Among the stream candidates, we note in particular two clusters in the Galactic northern hemisphere, shown in further detail in Figure~\ref{fig:NorthClusters}. The first cluster, which we dub the {\it Raritan}, is located in the upper right quadrant of Figure~\ref{fig:streamatlas}, and contains 14 \vmc{} candidates (with total significance 44.28) and 9 \vma{} candidates (with total significance 35.25), of which 6 overlap between both methods. The second cluster, which we dub the {\it Passaic}, is located in the upper left quadrant of Figure~\ref{fig:streamatlas}, and contains 4 \vmc{} candidates (with total significance 28.83) and 3 \vma{} candidates (with total significance 17.91), of which 3 overlap between both methods.

Both clusters are built out of individual streams identified using the algorithm described in the beginning of Section~\ref{sec:stream_candidates}. After this initial identification, additional by-eye grouping was performed, combining stream candidates into a larger object that appears to share similar bulk properties across stream components. Our loose criteria for this step was to discard nearby stream candidates that may overlap in position but were far apart in proper motion. Stream candidates that were overlapping in both position and proper motion with the rest of the cluster must also have a distribution of proper motion across the stream's position that broadly aligns with that of the rest of the cluster. Streams that did not follow this pattern were not included even if they overlapped in position and proper motion.

\begin{figure*}
\centering
    \subfigure{\includegraphics[width=0.79\textwidth, valign=t]{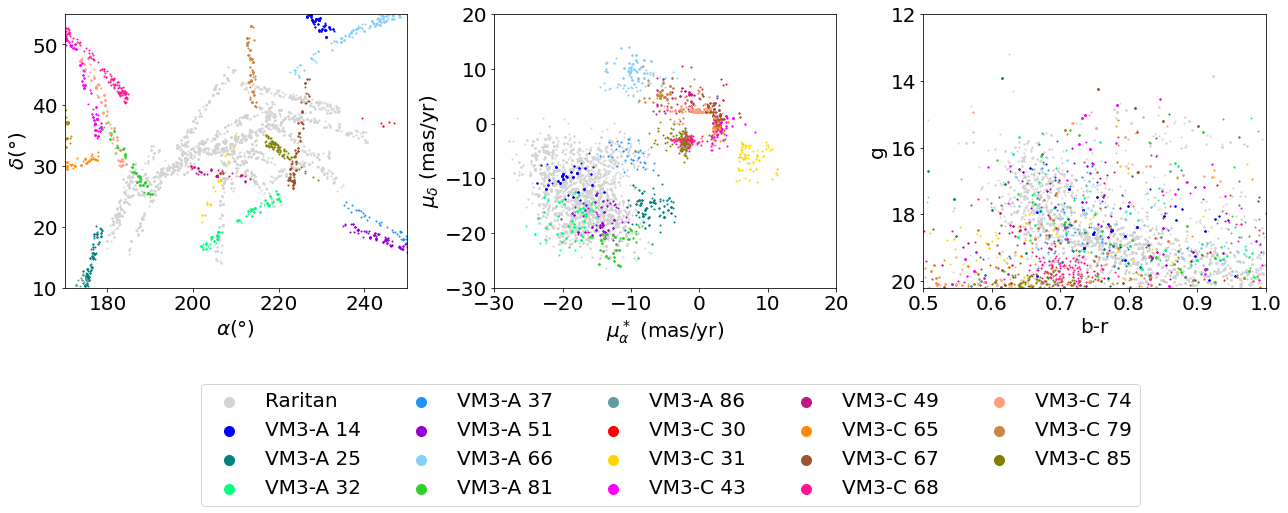}}    
    \subfigure{\includegraphics[width=0.79\textwidth, valign=t]{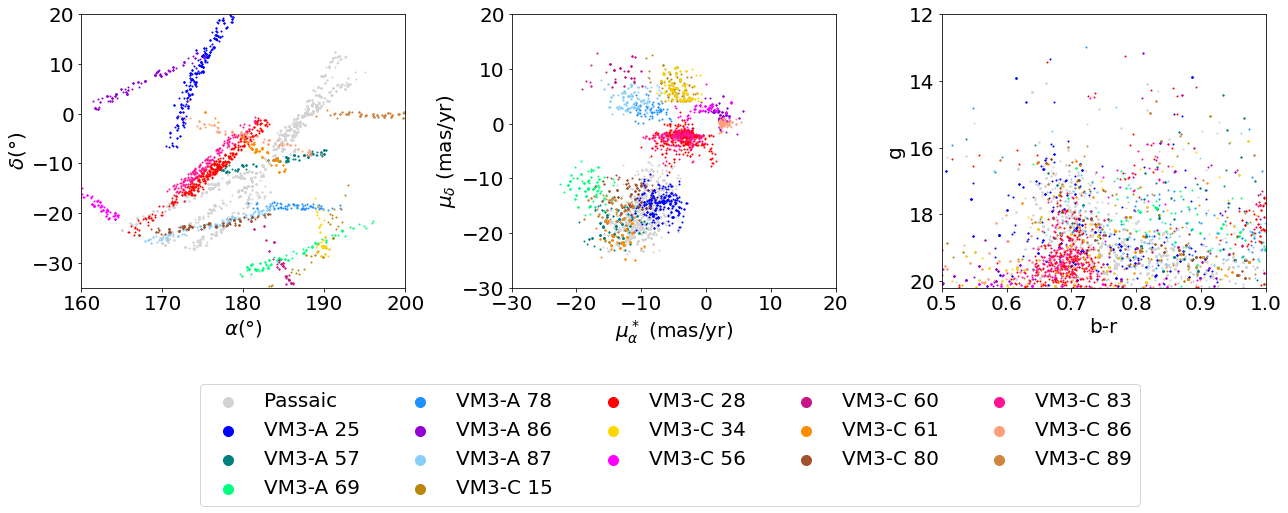}}
    \caption{Candidates, not corresponding to any known stream, in the vicinity of \textit{Raritan} (top) and \textit{Passaic} (bottom) respectively, that were not clustered to these objects.}
    \label{fig:NorthClusters_background}
\end{figure*}

Table~\ref{tab:cluster_contents} lists the stream candidates combined into each cluster using these rules of thumb, and Fig.~\ref{fig:NorthClusters_background} shows the candidates in the vicinity of the clusters that (following the reasoning above) were not merged into them. Most of these excluded stream candidates are clearly separated in position space and/or proper motion from the clusters in question. A few borderline cases merit further discussion:
\begin{itemize}
\item \vma{} 25: while its proper motion is close to the proper motions of the other candidates merged to \textit{Raritan} in that particular region of position space, we exclude it since it has a small but noticeable gap  in position space and proper motion space with the rest of the cluster. 
\item \vma{} 32: although this stream canddiate is overlapping with {\it Raritan} in proper motion space, it is excluded as it is well-separated from the bulk of the cluster in position space. 

\item \vmc{} 15 and \vma{} 57: while these are consistent with {\it Passaic} in proper motion and overlap in position, they are excluded because they do not follow the general direction of the cluster.

\end{itemize}

The position space plots of {\it Raritan} and {\it Passaic} are clearly not consistent with typical cold globular cluster streams.
They could instead be showing signs of multiple tidal tails and other deviations from a single thin filament (e.g., bifurcation or fanning), which may develop as a result of eccentric progenitor orbits \citep{2007ApJ...659.1212M,2008ApJ...689..936J,2015MNRAS.446.3100H,2015ApJ...799...28P,2025ApJ...979...75G}, chaotic orbits \citep{2016MNRAS.455.1079P,2020MNRAS.492.4398M}, or the stream overlapping multiple orbital resonances \citep{2015MNRAS.452..301F,2021MNRAS.501.1791Y,2023ApJ...954..215Y}. Further analysis of the orbits is necessary to confirm which (if any) of these mechanisms are at work here; at the present, the behavior of this object in position-space is consistent with a real stream. Additionally, the identification of multiple clusters within the same stream object can be compared to the \vm{} algorithm identifying fragments of Sagittarius (see Figure~\ref{fig:known_streams2}) rather than the entire stream. Sagittarius is simply too wide to be picked up in one piece by our search that is focused on narrow streams. A similar interaction may be at work here as well, i.e., we could be seeing fragments of a larger dwarf galaxy stream in {\it Raritan} and/or {\it Passaic}.

\section{Discussion and conclusions}
\label{sec:discussion_conclusions}
In this work, we have performed the first-ever search for stellar streams using the anomaly detection method \cathode{}. Originally developed to search for new physics in particle collider data, the \cathode{} method is completely general and can be applied to search for localized overdensities in any data. By combining \cathode{} with the \vm{} algorithm to focus on stream-like overdensities, we have produced a robust catalog of high significance stream candidates that we have dubbed \vmc{}. As further (minor) optimizations have been made to the steps of \vm{}, the previous catalog of stream candidates produced in \cite{Shih:2023jfv} using the closely related \anode{} anomaly detection method is also updated in this work. This we have dubbed \vma{}.

The full-sky scan using the unlabelled \textit{Gaia} DR2 re-discovers several known streams with high significance, and a considerable number of new stream candidates. Overall, \vmc{} results in 95 high-significance streams while \vma{} results in 87. Using the \textit{Galaxia} mock stellar catalog, the corresponding false positive rates are estimated to be 12.0\% and 12.8\% respectively. After taking into account Poisson statistics, the 95\% lower limit on the number of true stream candidates found by \vmc{} (\vma) is 84 (76).

Comparing the \vmc{} and \vma{} catalogs, we find that
\cathode{} produces a higher purity stream than \anode{} when tested on the GD-1 stream, using labels from an existing stellar catalog. In addition, where \vmc{} and \vma{} overlap, we find that \vmc{} generally increases the significances of stream candidates relative to \vma.

Overall, \vmc{} and \vma{} overlap on the rediscovery of 10 known streams (GD-1, Gaia-1, Jhelum, Fjorm, Gaia-8, Leiptr, Ylgr, ACS, Lethe and Sagittarius fragments). \vmc{} additionally finds Cetus-Palca; possibly a tidal stream associated to M68, the progenitor of Fjorm; and possibly part of the Triangulum-Pisces stream. Meanwhile, \vma{} finds Gaia-9 and Gaia-6, including a possible extension to Gaia-6 beyond what was reported in \cite{2019ApJ...872..152I}.

Beyond previously-known streams, \vmc{} and \vma{} combined discover a total of 25 stream candidates with high confidence that do not appear to correspond to previously discovered streams, and 15 of these are found by both \vmc{} and \vma. 

In addition, both \vmc{} and \vma{} find evidence of two separate clusters or agglomerations of stream-like objects in the Northern Galactic hemisphere. We have dubbed these the {\it Raritan} and the {\it Passaic}. Both are much broader and more complex than the other stream candidates that we find in VM3. We speculate that they could be remnants of a dwarf galaxy or other tidal debris. They could also possibly be GC streams showing signs of multiple tidal tails, fanning, or bifurcation. This behavior, if confirmed, would make these streams sensitive probes of the Galactic potential and the initial conditions of the progenitor objects.

While comparing the performance of \cathode{} to that of \anode{} in this work, it is also important to note that these two methods are not uncorrelated. Both use the same sideband training to estimate the background in the signal region. If this sideband training is flawed, it could introduce the same artifacts in both methods. Nevertheless, from tests on simulated LHC data \citep{Hallin:2021wme, Golling:2023yjq}, it appears that \cathode{} is more powerful at detecting anomalous events, even with the same sideband training. For this reason, finding a stream with high significance in both methods does increases the probability that it is a real stream rather than a false positive. In order to confirm whether the stream candidates are indeed real streams, they need to be further studied (for example, measuring orbits, radial velocities, and metallicities either through cross-matching with existing surveys or dedicated observations). To facilitate such studies by the research community, the stream data will be made public.

In addition to follow-up observations of stream candidates, we see several avenues for future work. First of all, in order to be comparable to VM2, this work used data from DR2 rather than the newer DR3. The next step will be to apply \vmc{} to DR3 or the upcoming DR4, which contains more stars and improved measurements compared to DR2. This would also give an independent cross check in the sense that the density estimator from \cite{Shih:2023jfv} could not be re-used, but would have to be retrained. Due to the computational cost, it was not possible in this work to train and ensemble several density estimators (as was done in \cite{Hallin:2021wme}). If future model improvements lower the required amount of compute, such an ensembling would be valuable in order to improve the stability and trustworthiness of the density estimator.

Furthermore, the only handle we have to rank the stream candidates so far is the significance. This means that we believe that a high significance candidate is to be more likely to be real than one with low significance. However, we do find matches to known streams where the stream candidate has ended up rather far down the significance list, most notably Lethe. Because of this, we need to develop additional handles to rank the stream candidates. 

Finally, we also see great value in applying other anomaly detection techniques to \textit{Gaia} data \citep{Pettee:2023zra,Sengupta:2024ezl}, as they can provide both confirmation of previously identified stream candidates and complementarity by finding additional candidates. 

\section{Data availability}
The \textit{Gaia} DR2 data upon which this work is based, is publicly available. The member stars of each stream candidate will be made public after the manuscript is accepted for publication.

\begin{acknowledgements}
The work of AH was supported in part by the Deutsche Forschungsgemeinschaft under Germany’s Excellence Strategy – EXC 2121  Quantum Universe – 390833306, and under PUNCH4NFDI – project number 460248186, and in part by the US Department of Energy under grant DE-SC0010008. MRB and DS are supported by DOE grant DOE-SC0010008.
The authors acknowledge the Office of Advanced Research Computing (OARC) at Rutgers, The State University of New Jersey for providing access to the Amarel cluster and associated research computing resources that have contributed to the results reported here. This research used resources of the National Energy Research
Scientific Computing Center (NERSC), a U.S. Department of Energy
Office of Science User Facility operated under Contract No. DEAC02-05CH11231.
This work has made use of data from the European Space Agency (ESA) mission \textit{Gaia} (\url{https://www.cosmos.esa.int/gaia}), processed by the \textit{Gaia} Data Processing and Analysis Consortium (DPAC, \url{https://www.cosmos.esa.int/web/gaia/dpac/consortium}). Funding for the DPAC has been provided by national institutions, in particular the institutions participating in the \textit{Gaia} Multilateral Agreement. 
This work was performed in part at the Aspen Center for Physics, which is supported by National Science Foundation grant PHY-2210452.

\end{acknowledgements}

\clearpage

\appendix
\onecolumn

\section{Previously known streams}
\label{sec:knownstreams}
\begin{figure}[h]
    \centering
    \subfigure{\includegraphics[width=0.79\textwidth, valign=t]{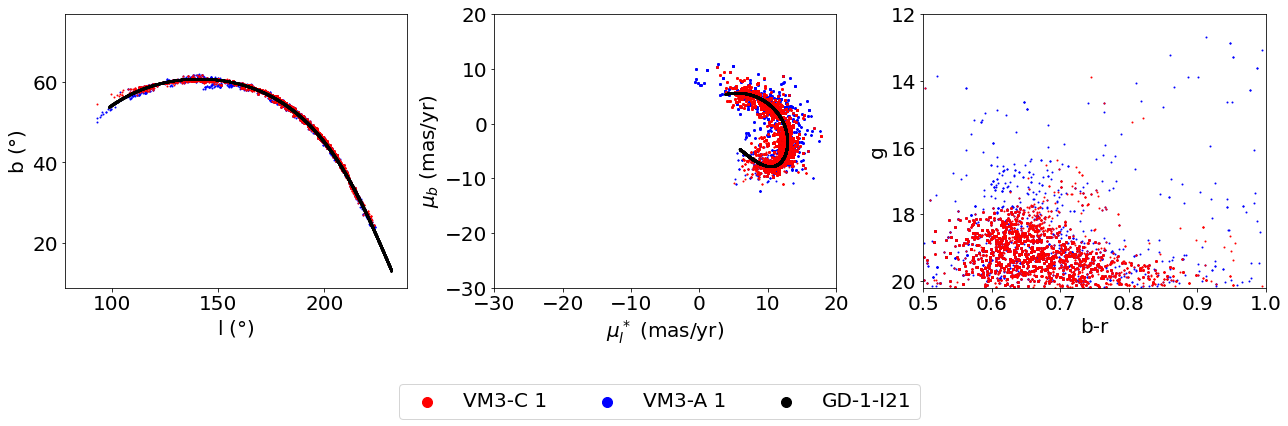}}
    \subfigure{\includegraphics[width=0.79\textwidth, valign=t]{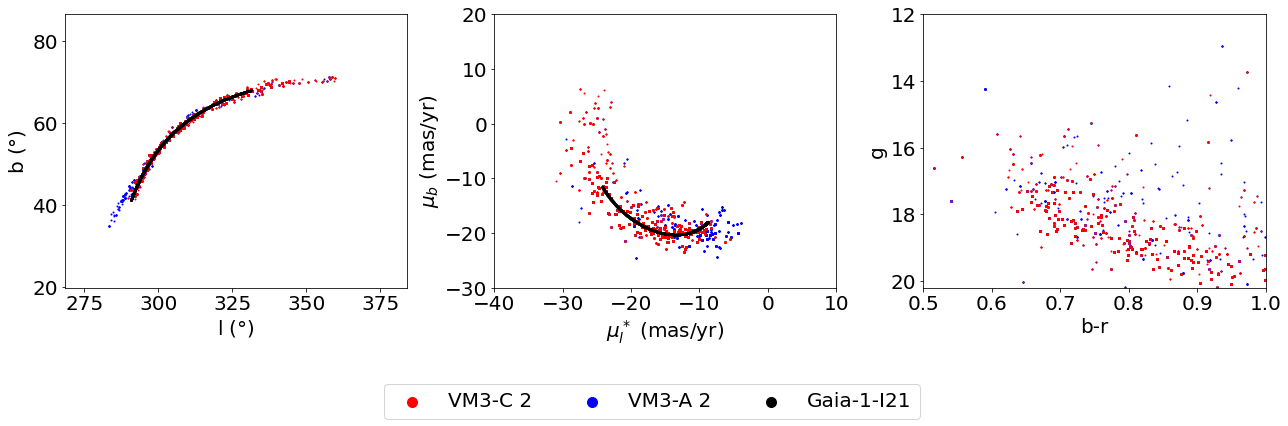}}
    \subfigure{\includegraphics[width=0.79\textwidth, valign=t]{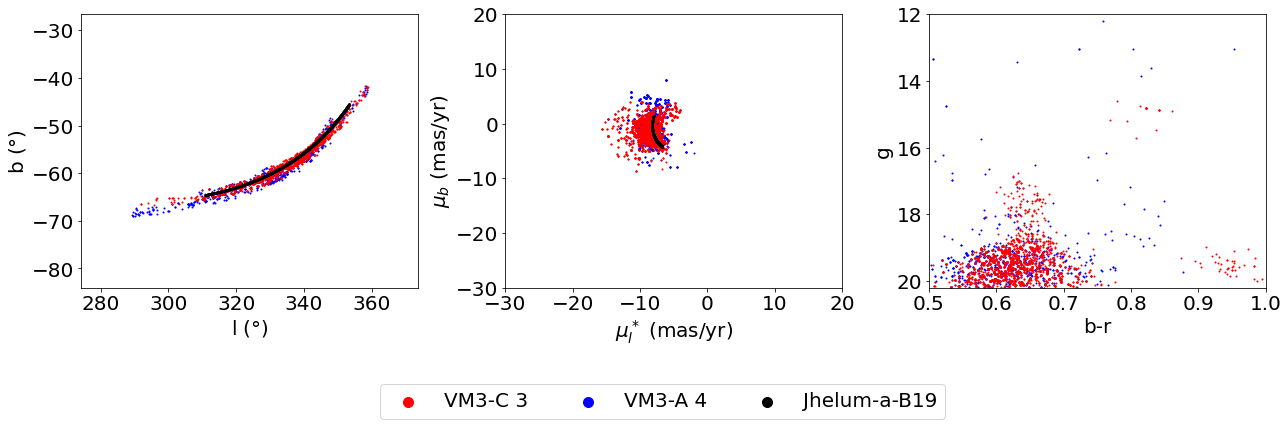}} 
    \subfigure{\includegraphics[width=0.79\textwidth, valign=t]{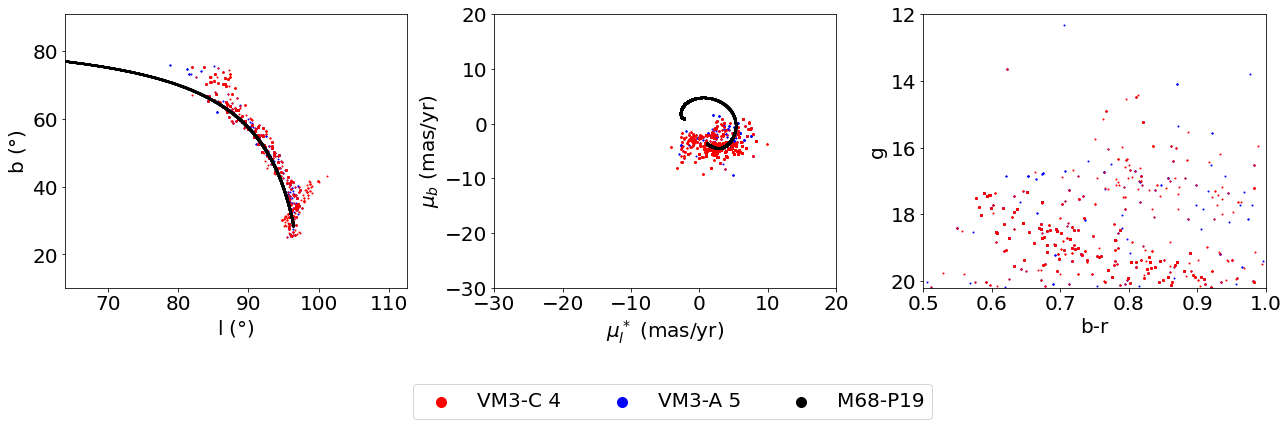}}
    \caption{Performance of \vmc{} and \vma{} on the following known streams: GD-1, Gaia-1, Jhelum, Fjorm. The \vmc{} stars are plotted in red and the \vma{} stars in blue. Overlaid in black are stream tracks from the \texttt{galstreams} stellar stream library.}
    \label{fig:known_streams1}
\end{figure}

\begin{figure}[h]
    \centering
    \subfigure{\includegraphics[width=0.79\textwidth, valign=t]{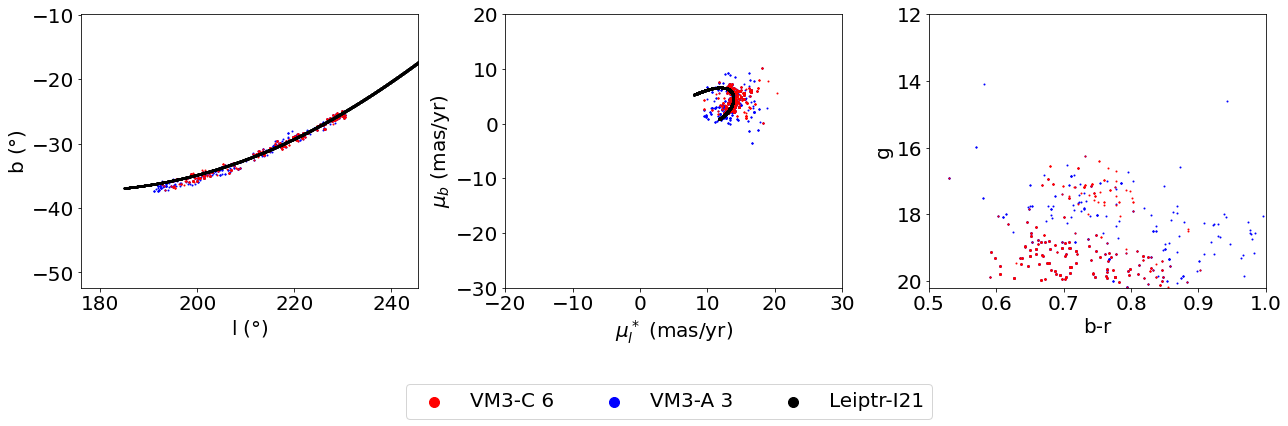}}
    \subfigure{\includegraphics[width=0.79\textwidth, valign=t]{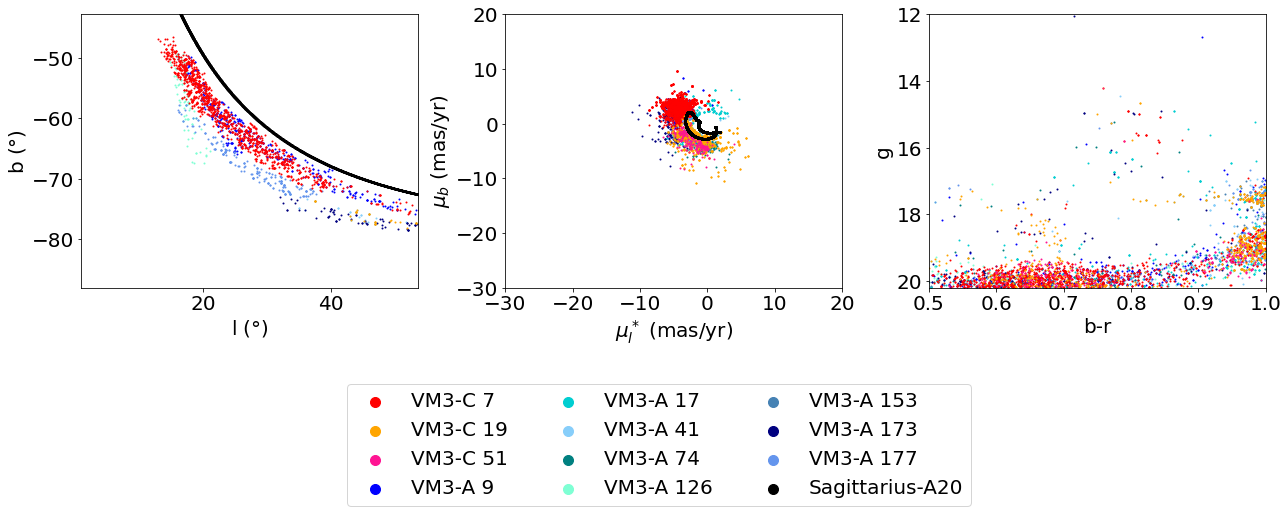}}
    \subfigure{\includegraphics[width=0.79\textwidth, valign=t]{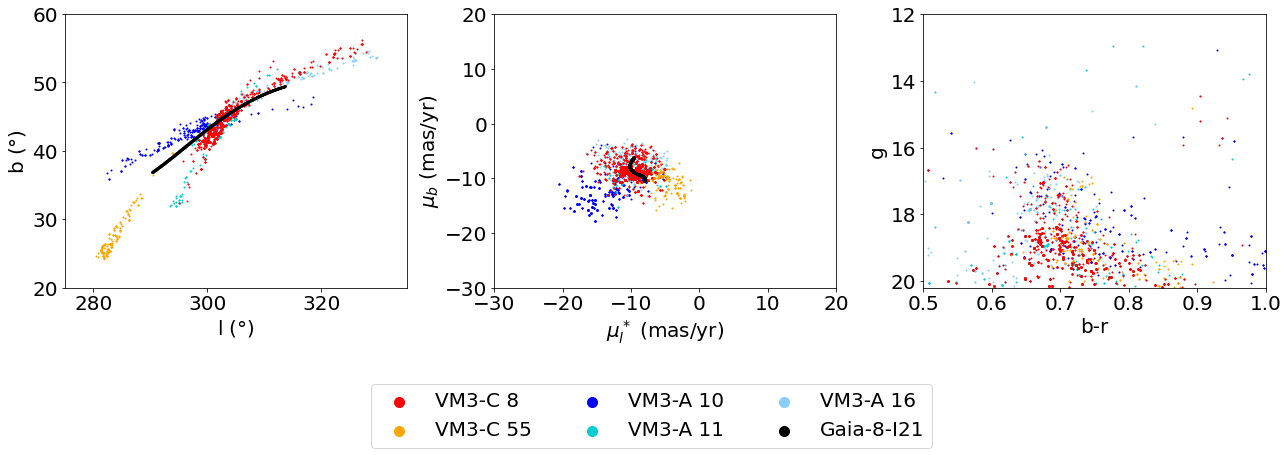}}
    \subfigure{\includegraphics[width=0.79\textwidth, valign=t]{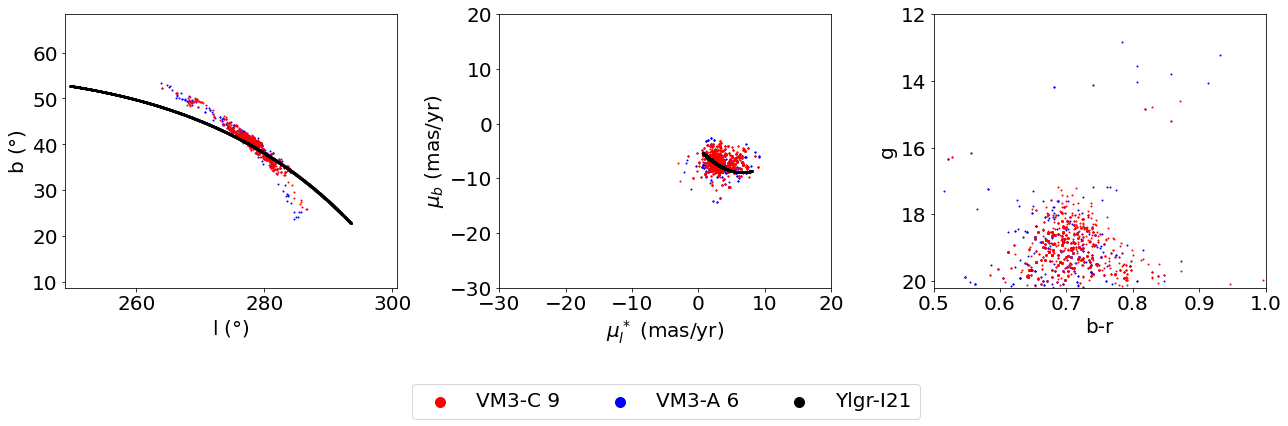}}
    \caption{Performance of \vmc{} and \vma{} on the following known streams: Leiptr, Sagittarius, Gaia-8 and Ylgr. The \vmc{} stars are plotted in red/pink/orange and the \vma{} stars in various shades of blue. Overlaid in black are stream tracks from the \texttt{galstreams} stellar stream library. VM3-C 55 is an extension of Gaia-8 compared to the track in \texttt{galstreams}, as previously found by \cite{ibata2023chartinggalacticaccelerationfield}.}
    \label{fig:known_streams2}
\end{figure}

\begin{figure}[h]
    \centering
    \subfigure{\includegraphics[width=0.79\textwidth, valign=t]{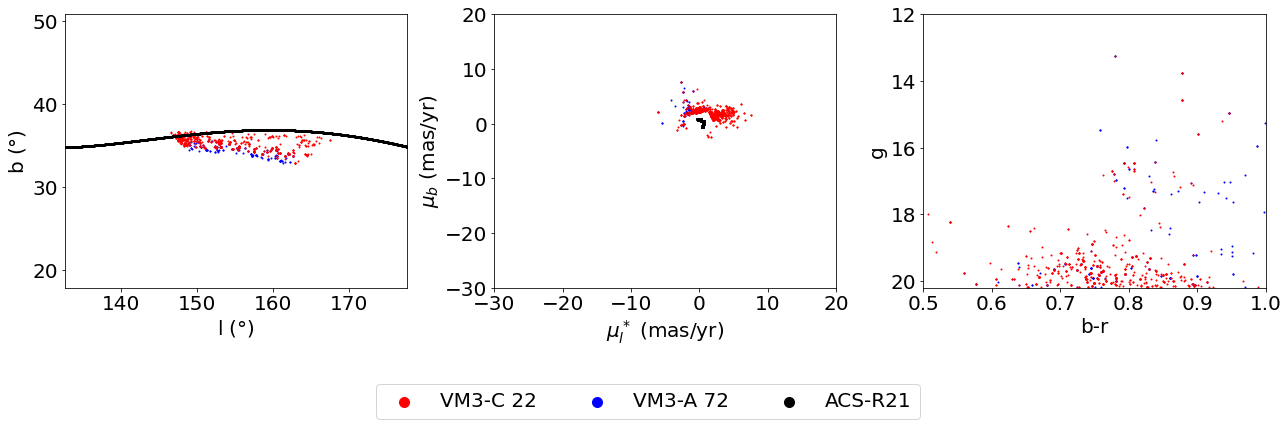}}
    \subfigure{\includegraphics[width=0.79\textwidth, valign=t]{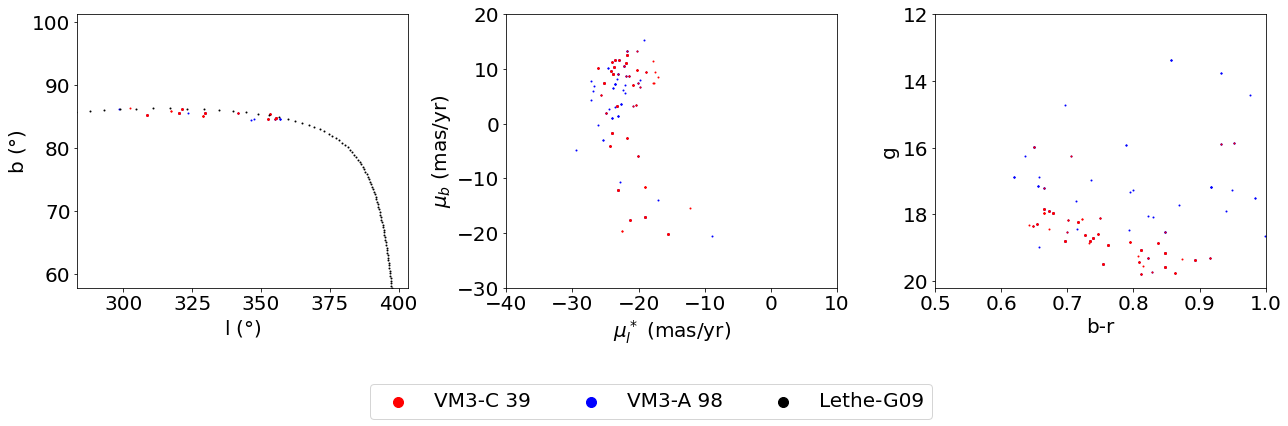}}
    \caption{Performance of \vmc{} and \vma{} on the previously known streams ACS and Lethe. The \vmc{} stars are plotted in red and the \vma{} stars in blue. Overlaid in black are stream tracks from the \texttt{galstreams} stellar stream library. Note that \texttt{galstreams} does not contain any proper motion data for Lethe.}
    \label{fig:known_streams3}
\end{figure}

\begin{figure}[h]
    \centering
    \subfigure{\includegraphics[width=0.79\textwidth, valign=t]{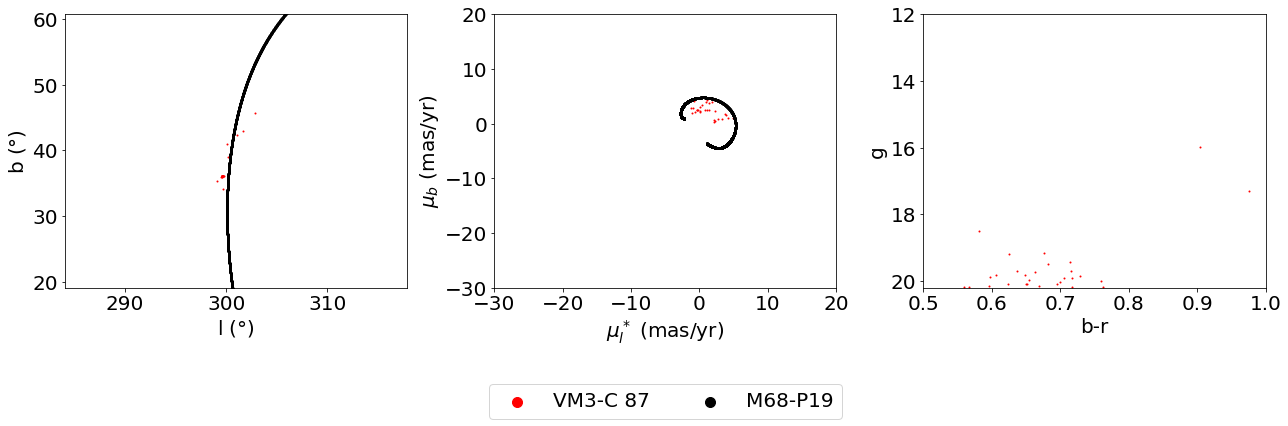}}
    \subfigure{\includegraphics[width=0.79\textwidth, valign=t]{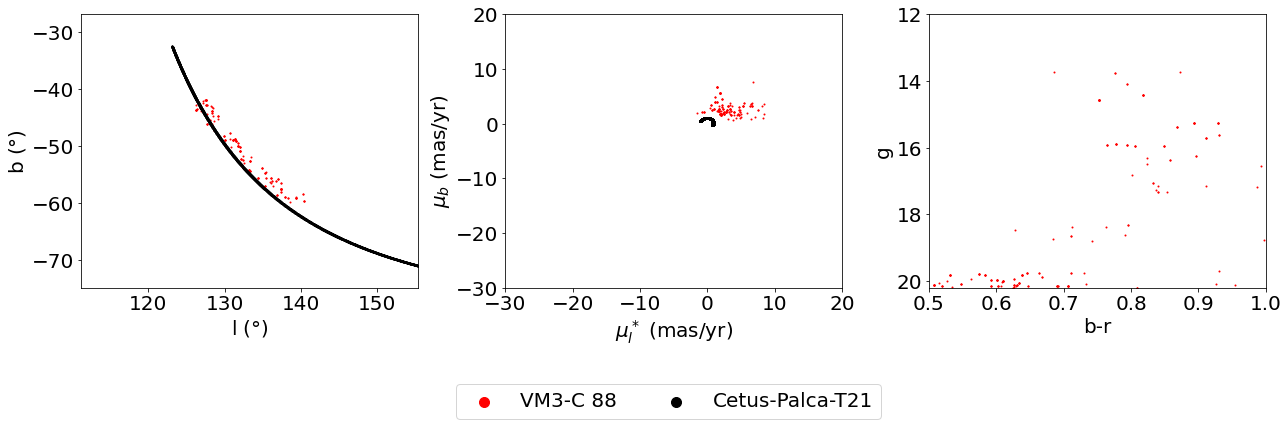}}
    \caption{Performance of \vmc{} on the previously known streams 
    M68 and Cetus-Palca. The \vmc{} stars are plotted in red, overlaid with the stream tracks from the \texttt{galstreams} stellar stream library in black. \vma{} did not detect these streams.}
    \label{fig:known_streams4}
\end{figure}

\begin{figure}[h]
    \centering
    \subfigure{\includegraphics[width=0.79\textwidth, valign=t]{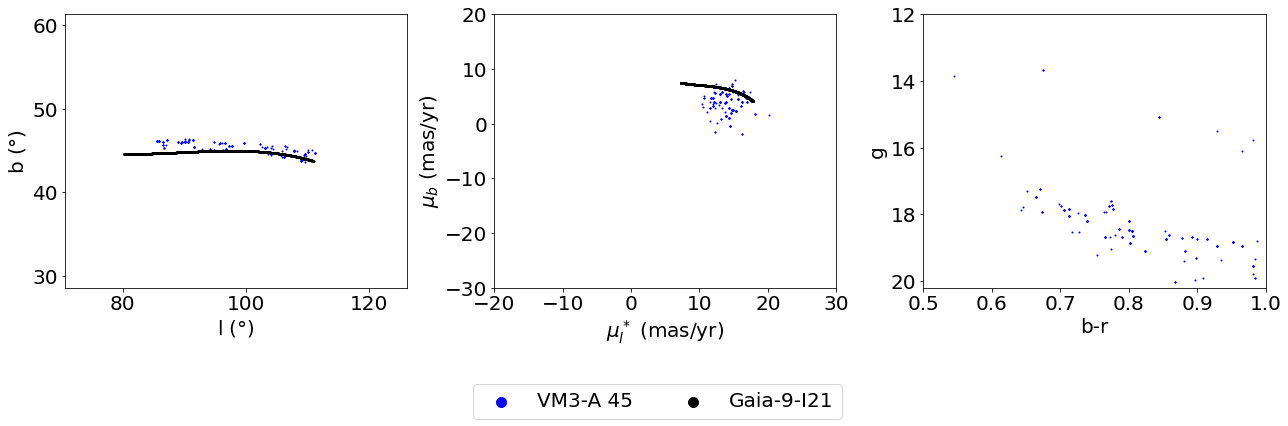}}
    \subfigure{\includegraphics[width=0.79\textwidth, valign=t]{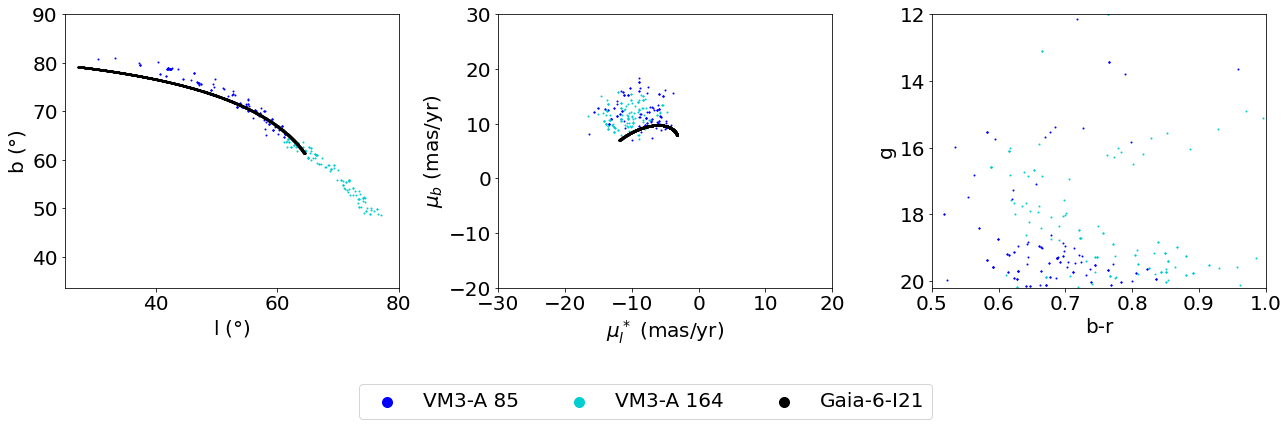}}
    \caption{Performance of \vma{} on the previously known streams Gaia-9 and Gaia-6. The \vma{} stars are plotted in blue, overlaid with the stream tracks from the \texttt{galstreams} stellar stream library in black. \vmc{} did not detect these streams.}
    \label{fig:known_streams5}
\end{figure}

Figures~\ref{fig:known_streams1}, \ref{fig:known_streams2}, \ref{fig:known_streams3}, \ref{fig:known_streams4}, and \ref{fig:known_streams5} show \vmc{} and \vma{} stream candidates that correspond to known streams, visualized together with stream tracks from the \texttt{galstreams} stellar stream library.

\begin{figure}[h]
    \centering
    \subfigure{\includegraphics[width=0.79\textwidth, valign=t]{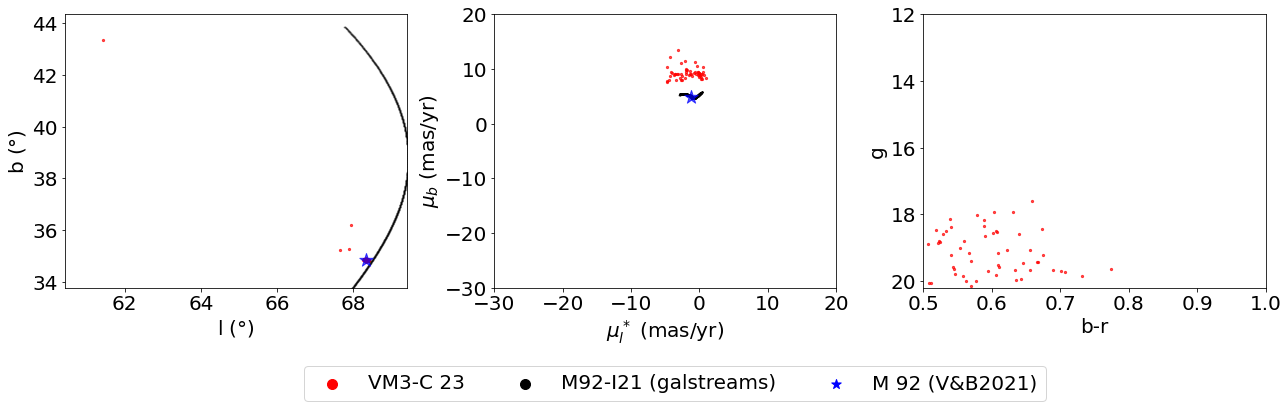}}
    \caption{Stream candidate 23 of \vmc{} shows a failure mode of the line-finding algorithm. The \vmc{} stars are plotted in red and the globular cluster M92 in blue~\citep{{10.1093/mnras/stab1475}} and black~(\texttt{galstreams}). See the text for further discussion.}
    \label{fig:no_stream}
\end{figure}

\clearpage

\section{New stream candidates}
\label{sec:newstreamcands}
\begin{figure}[h]
    \centering
    \subfigure{\includegraphics[width=0.79\textwidth, valign=t]{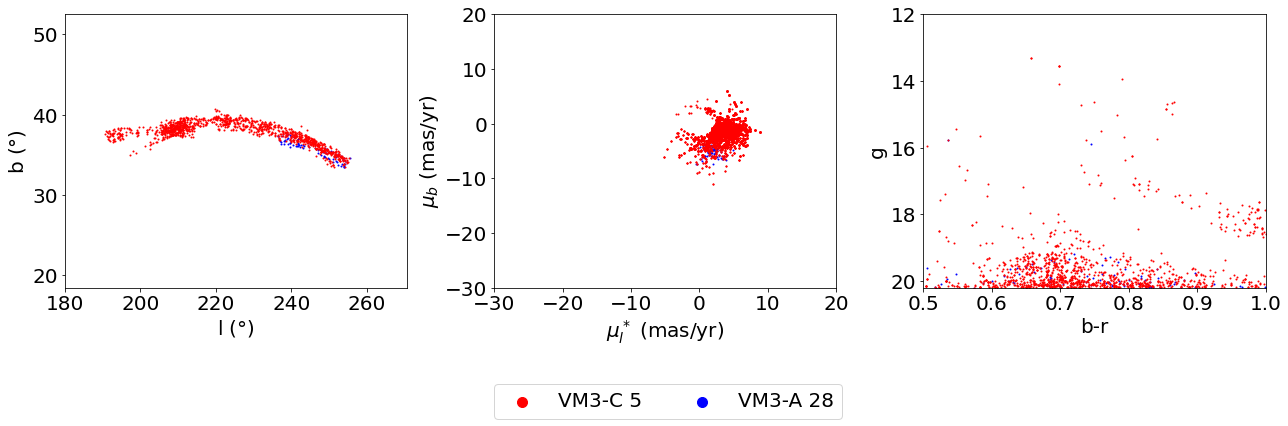}}
    \subfigure{\includegraphics[width=0.79\textwidth, valign=t]{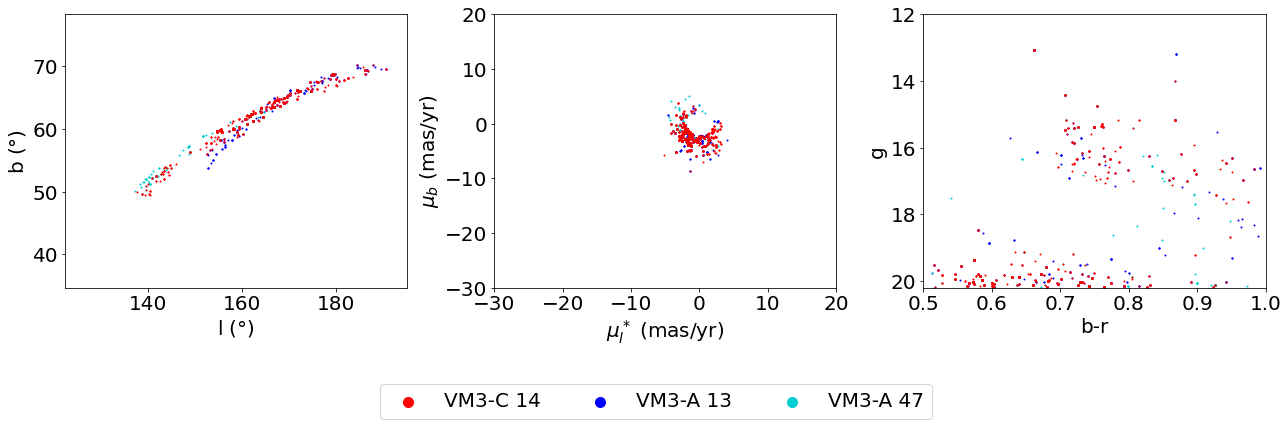}}
    \subfigure{\includegraphics[width=0.79\textwidth, valign=t]{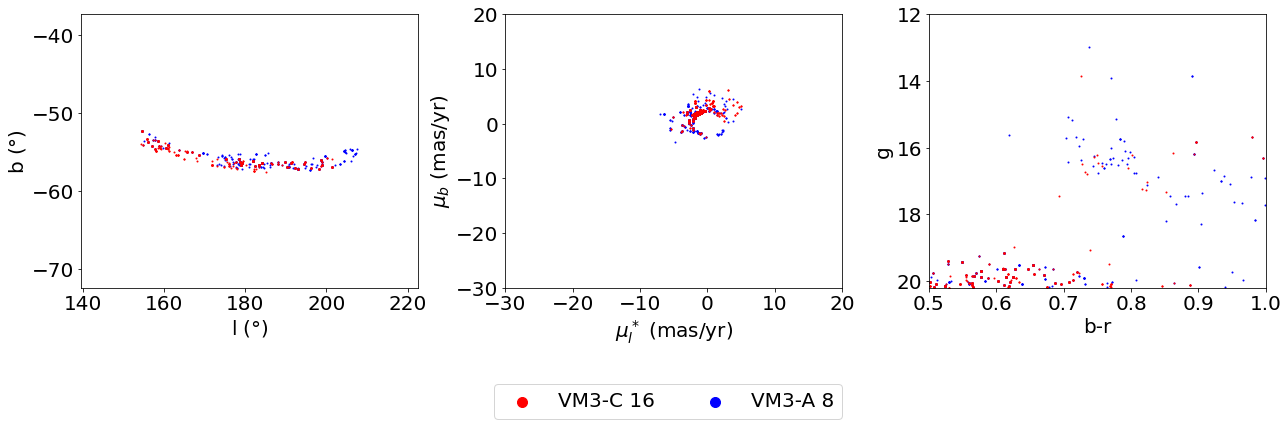}} 
    \subfigure{\includegraphics[width=0.79\textwidth, valign=t]{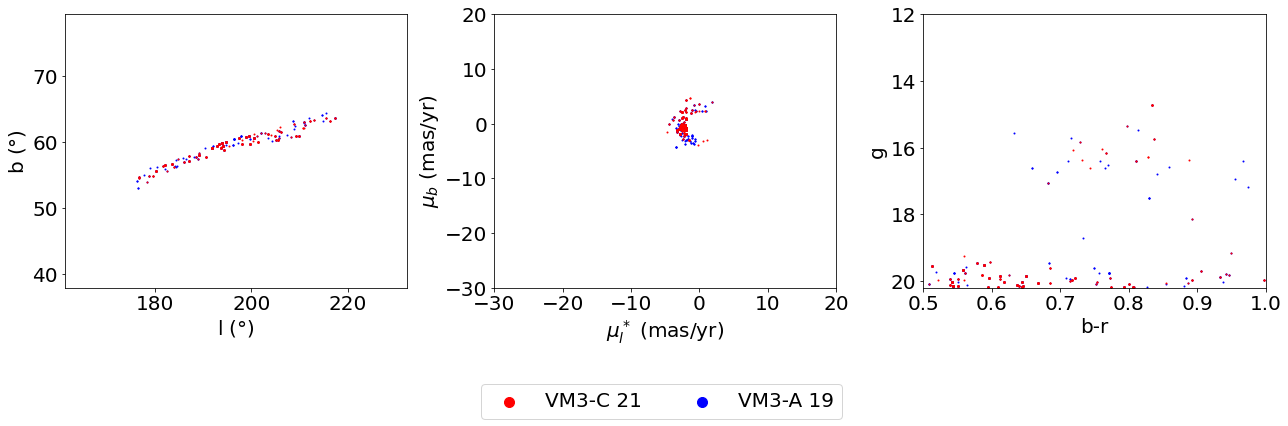}}
    \caption{New stream candidates identified in both \vmc{} and \vma{}. The \vmc{} stars are plotted in red/pink/orange and the \vma{} stars in various shades of blue..}
    \label{fig:new_streams1}
\end{figure}

\begin{figure}[h]
    \centering    
    \subfigure{\includegraphics[width=0.79\textwidth, valign=t]{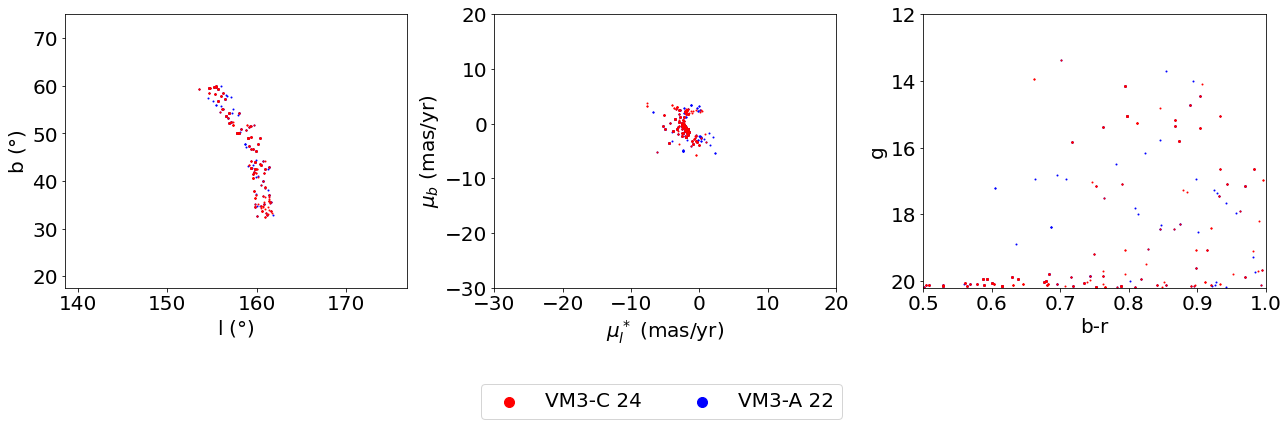}}
    \subfigure{\includegraphics[width=0.79\textwidth, valign=t]{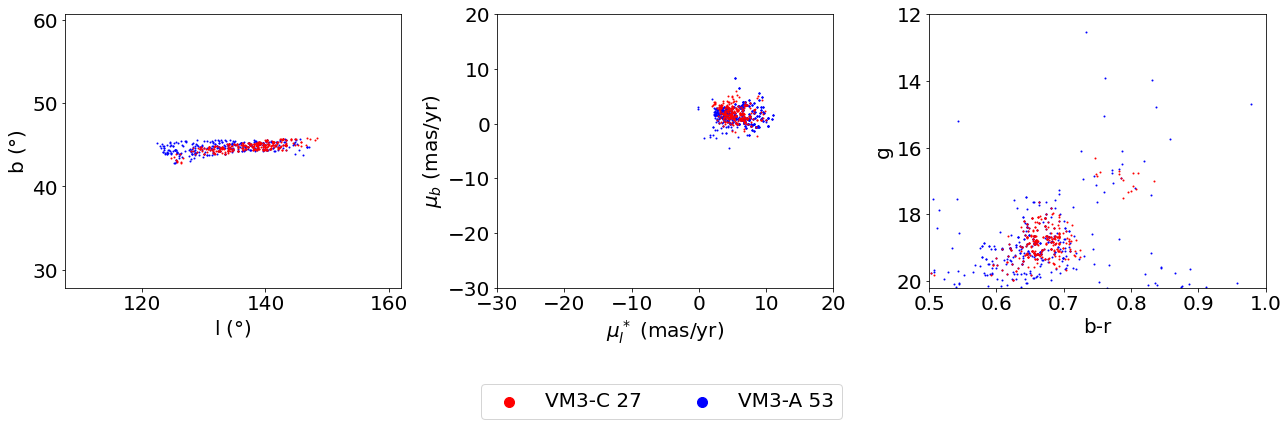}}
    \subfigure{\includegraphics[width=0.79\textwidth, valign=t]{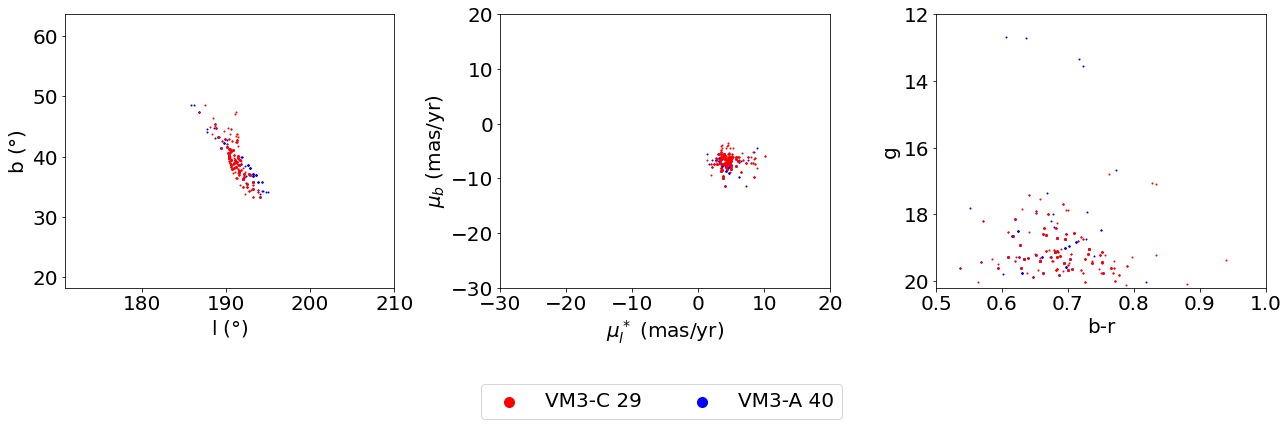}}
    \subfigure{\includegraphics[width=0.79\textwidth, valign=t]{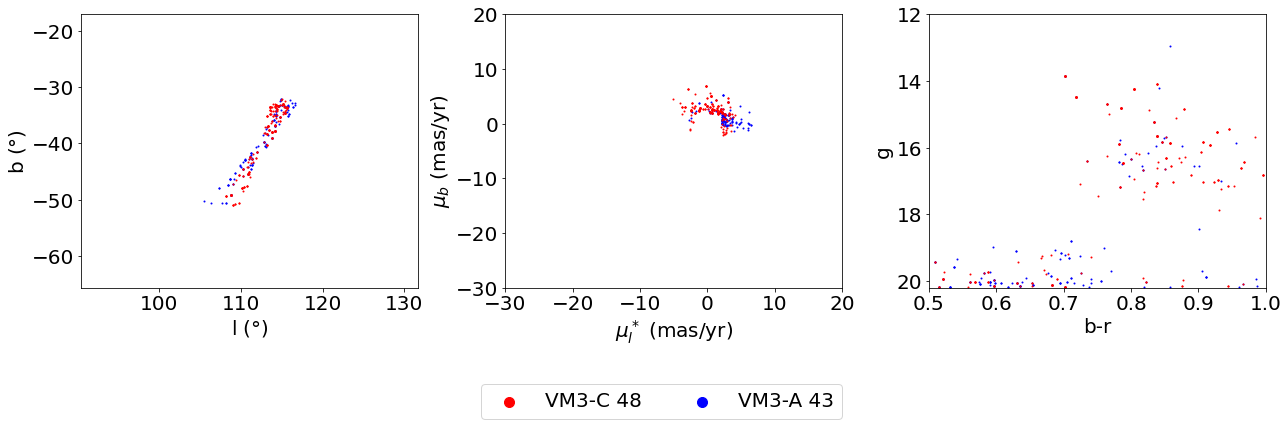}}
    \caption{New stream candidates identified in both \vmc{} and \vma{}. The \vmc{} stars are plotted in red and the \vma{} stars in blue.}
    \label{fig:new_streams2}
\end{figure}

\begin{figure}[h]
    \centering
    \subfigure{\includegraphics[width=0.79\textwidth, valign=t]{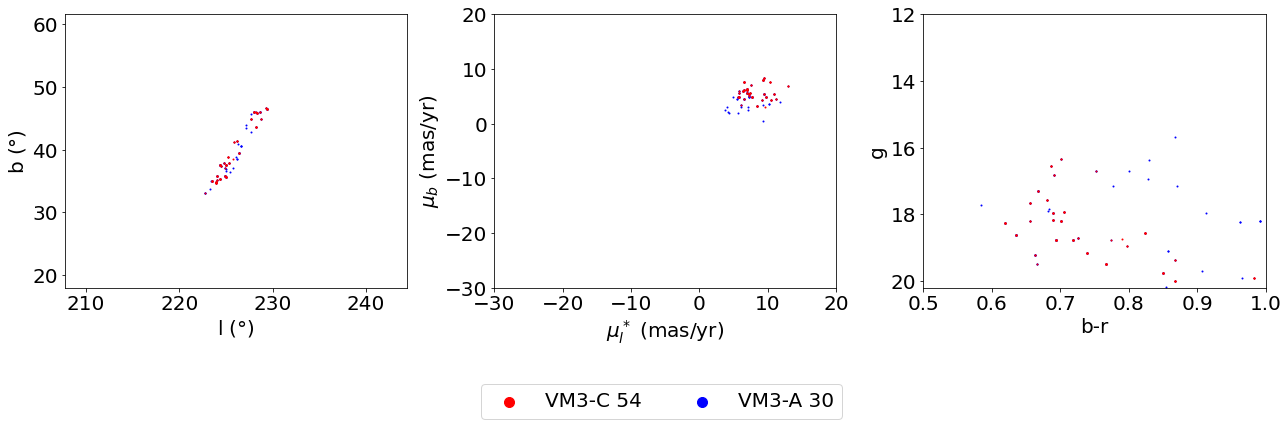}}
    \subfigure{\includegraphics[width=0.79\textwidth, valign=t]{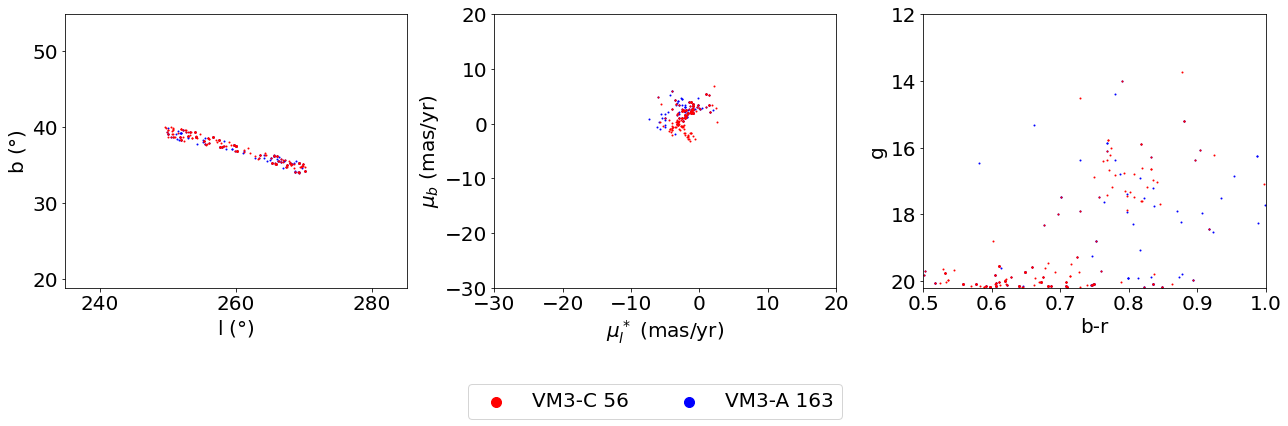}}
    \subfigure{\includegraphics[width=0.79\textwidth, valign=t]{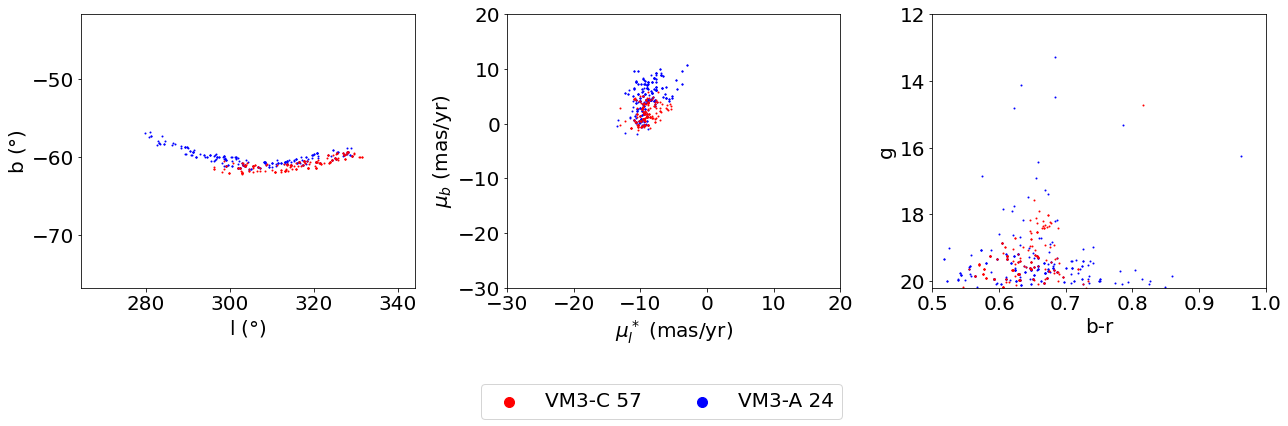}}
    \subfigure{\includegraphics[width=0.79\textwidth, valign=t]{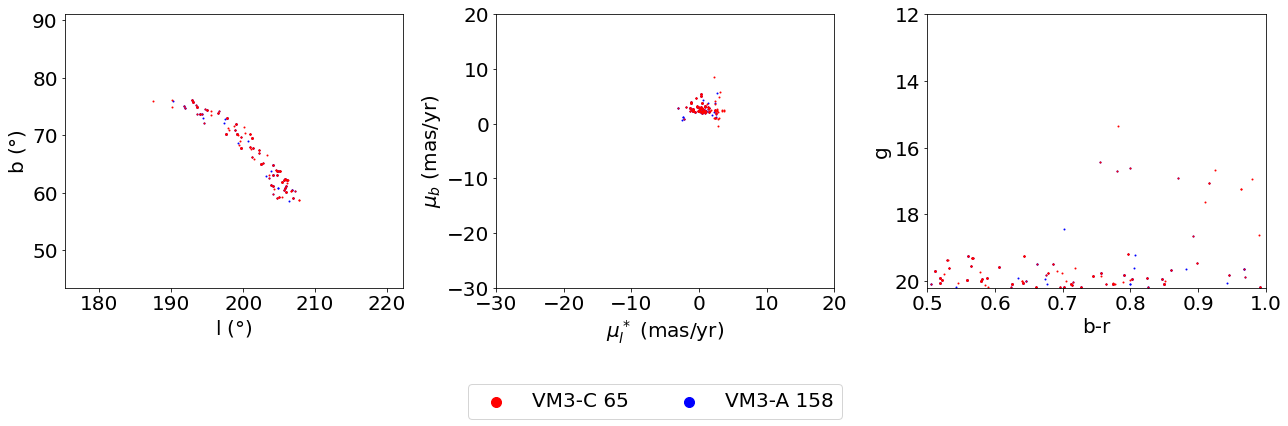}}
    \caption{New stream candidates identified in both \vmc{} and \vma{}. The \vmc{} stars are plotted in red and the \vma{} stars in blue.}
    \label{fig:new_streams3}
\end{figure}

\begin{figure}[h]
    \centering
    \subfigure{\includegraphics[width=0.79\textwidth, valign=t]{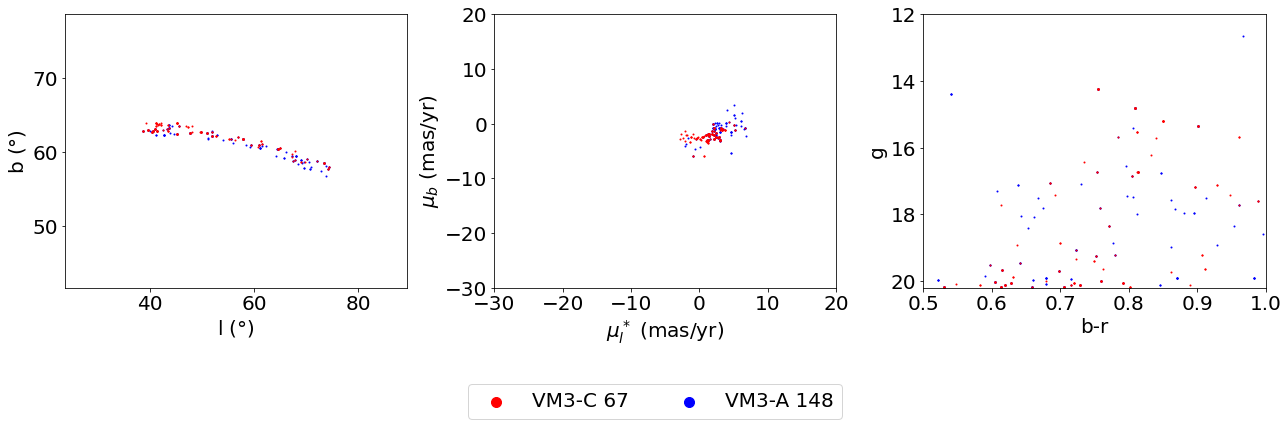}}
    \subfigure{\includegraphics[width=0.79\textwidth, valign=t]{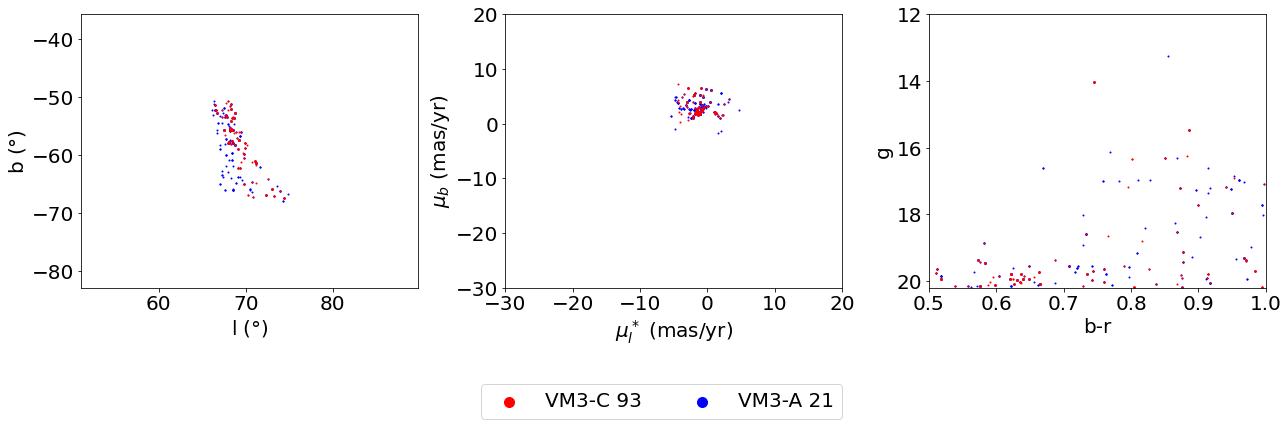}}
    \caption{New stream candidates identified in both \vmc{} and \vma{}. The \vmc{} stars are plotted in red and the \vma{} stars in blue.}
    \label{fig:new_streams4}
\end{figure}

\begin{figure}[h]
    \centering
    \subfigure{\includegraphics[width=0.79\textwidth, valign=t]{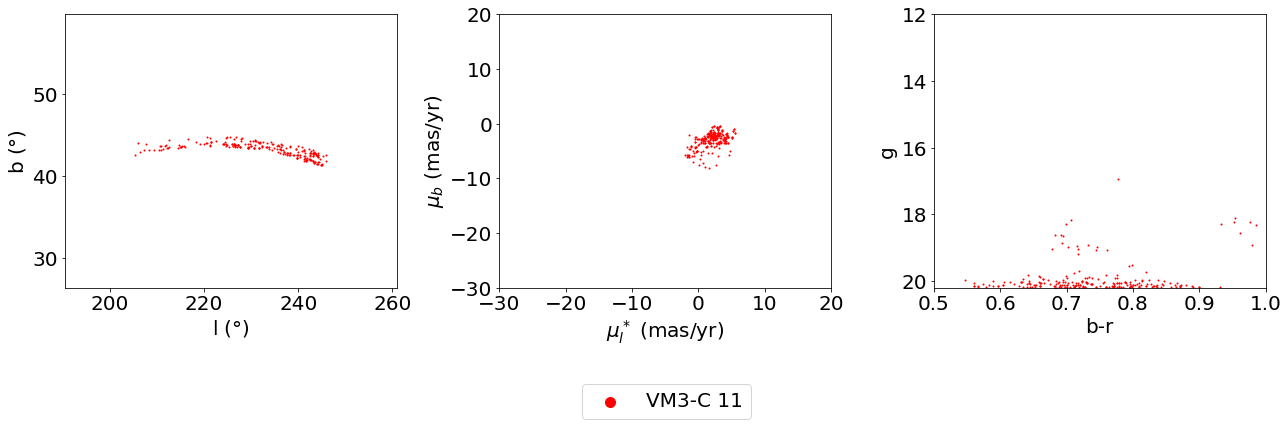}} 
    \subfigure{\includegraphics[width=0.79\textwidth, valign=t]{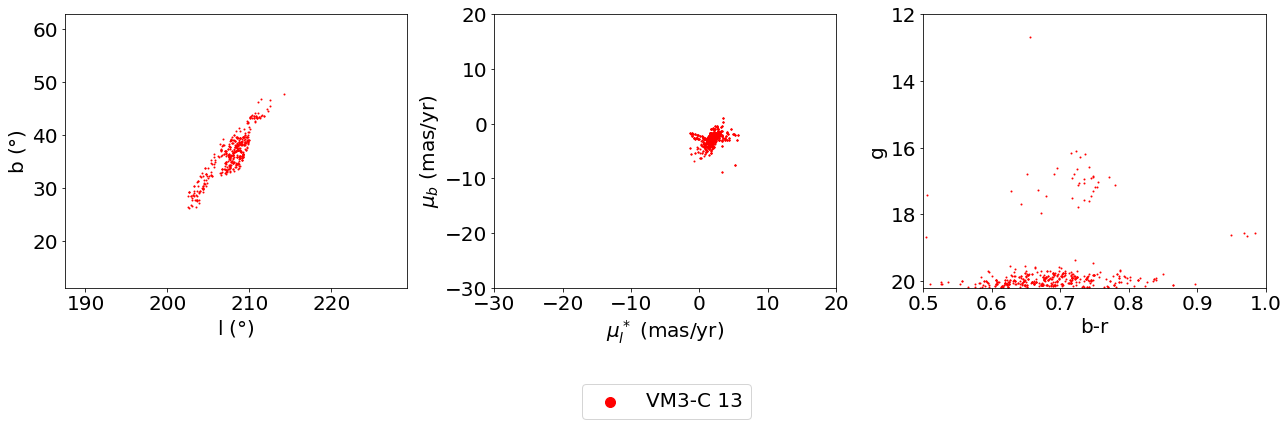}}
    \subfigure{\includegraphics[width=0.79\textwidth, valign=t]{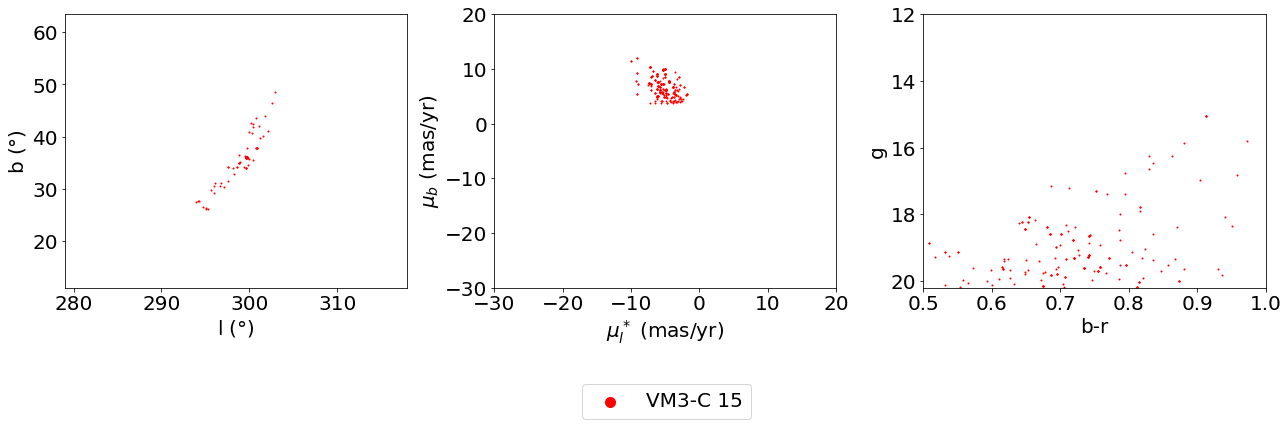}}
    \subfigure{\includegraphics[width=0.79\textwidth, valign=t]{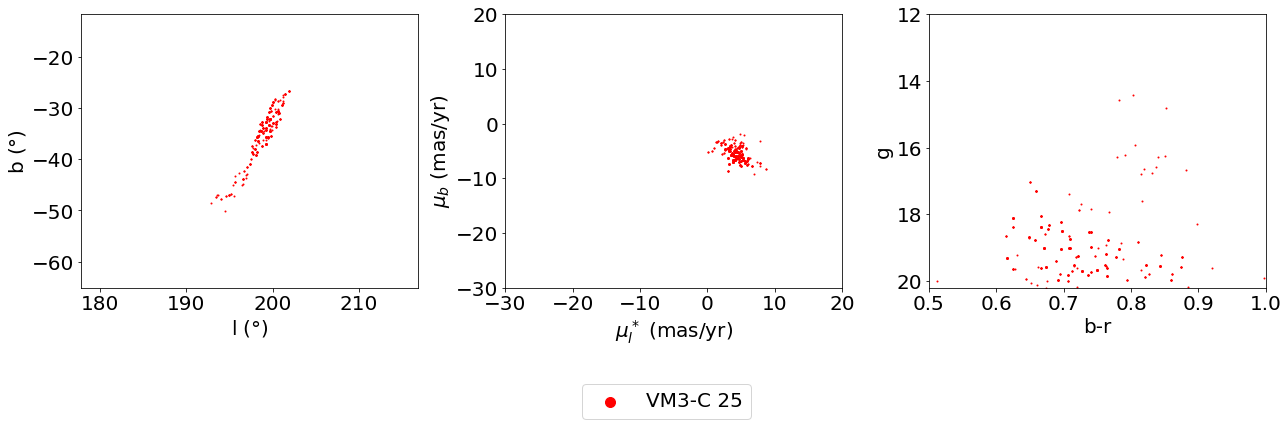}}
    \caption{New stream candidates identified in \vmc{} but not in \vma{}.}
    \label{fig:new_streams5}
\end{figure}

\begin{figure}[h]
    \centering
    \subfigure{\includegraphics[width=0.79\textwidth, valign=t]{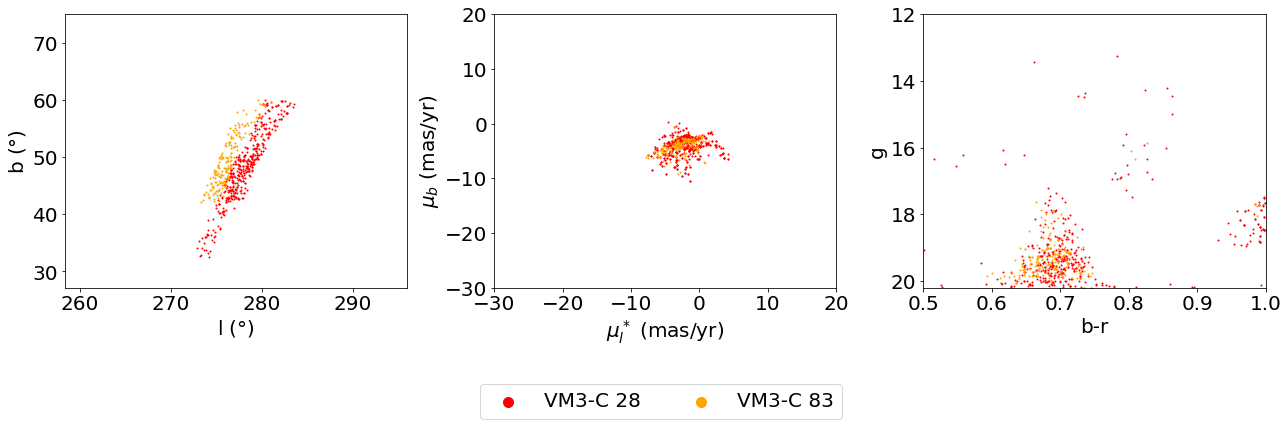}}
    \caption{New stream candidates identified in \vmc{} but not in \vma{}.}
    \label{fig:new_streams6}
\end{figure}

\begin{figure}[h]
    \centering
    \subfigure{\includegraphics[width=0.79\textwidth, valign=t]{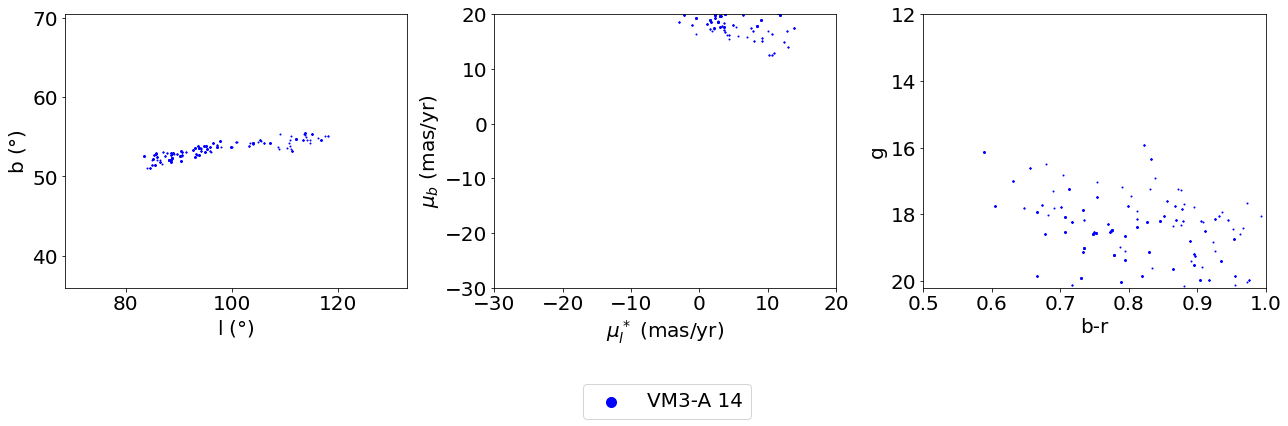}}
    \subfigure{\includegraphics[width=0.79\textwidth, valign=t]{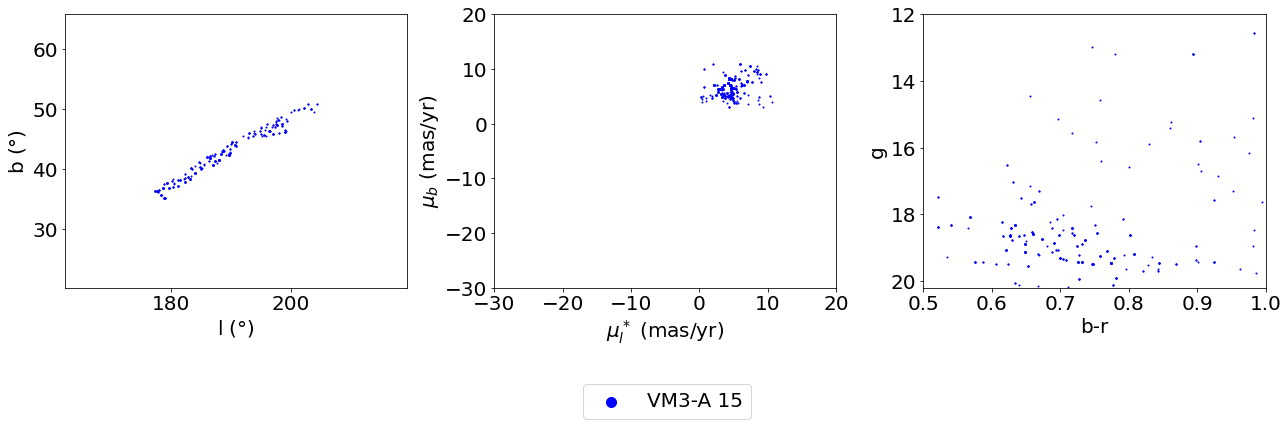}}
    \subfigure{\includegraphics[width=0.79\textwidth, valign=t]{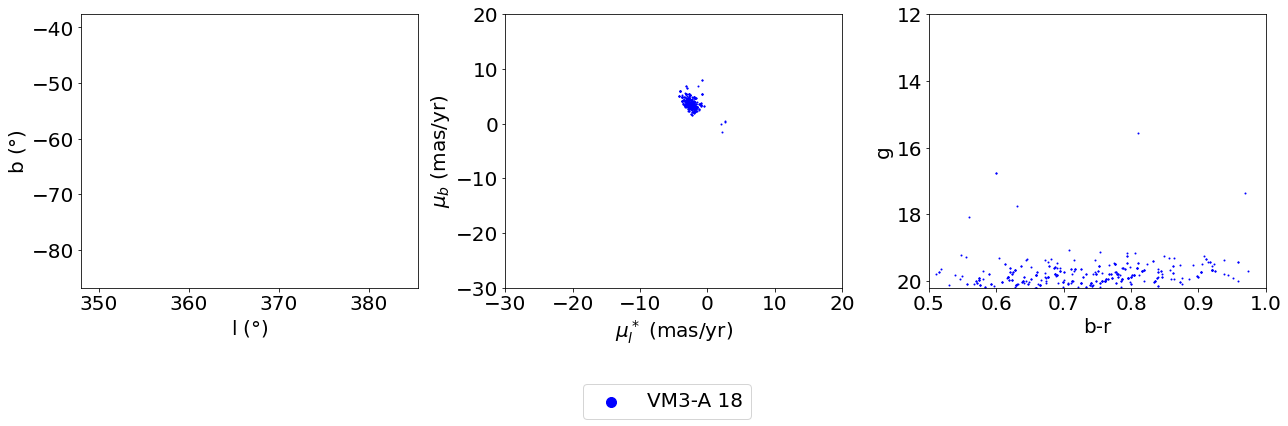}}
    \subfigure{\includegraphics[width=0.79\textwidth, valign=t]{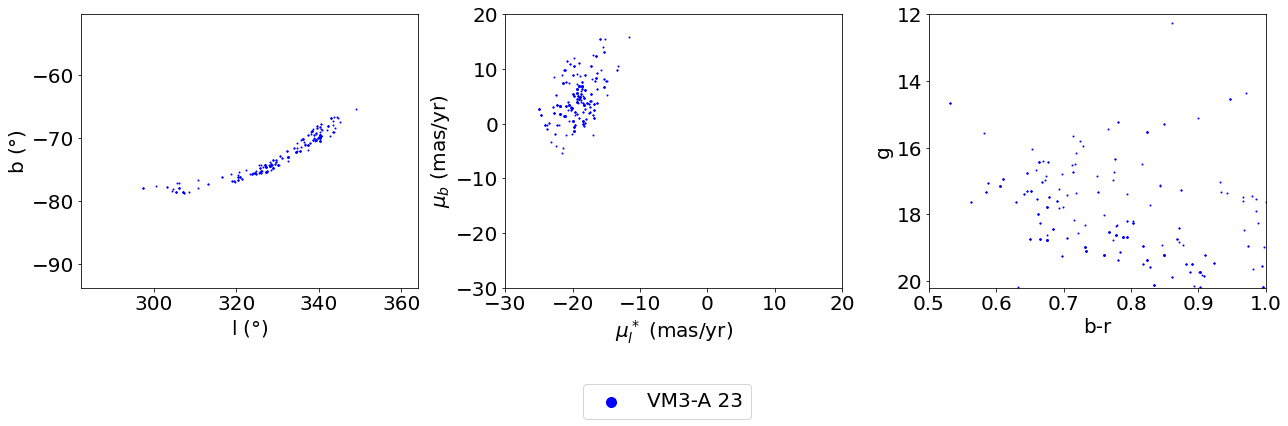}}
    \caption{New stream candidates identified in \vma{} but not in \vmc{}.}
    \label{fig:new_streams7}
\end{figure}

\begin{figure}[h]
    \centering
    \subfigure{\includegraphics[width=0.79\textwidth, valign=t]{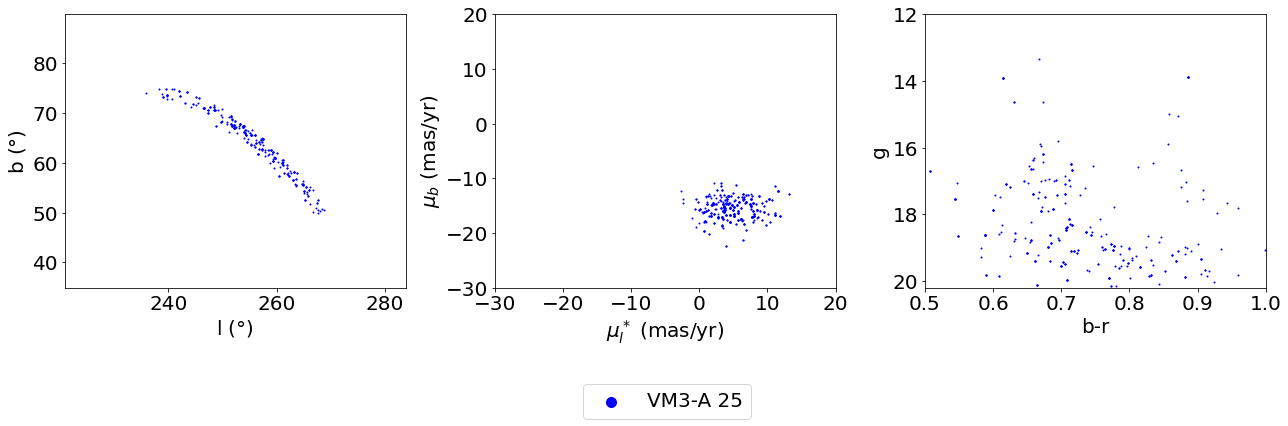}}
    \subfigure{\includegraphics[width=0.79\textwidth, valign=t]{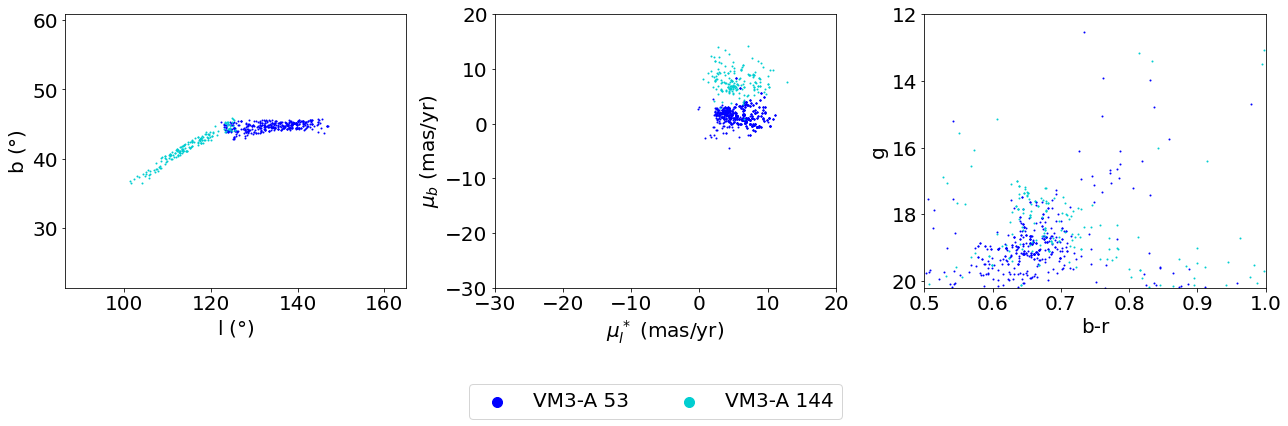}}
    \caption{New stream candidates identified in \vma{} but not in \vmc{}.}
    \label{fig:new_streams8}
\end{figure}

Figures~\ref{fig:new_streams1}, \ref{fig:new_streams2}, \ref{fig:new_streams3} and \ref{fig:new_streams4} show new stream candidates that have been found with \vmc{} and \vma{}. Figures \ref{fig:new_streams5} and \ref{fig:new_streams6} show new stream candidates that have been found only in \vmc, but not in \vma; whereas Figs.~\ref{fig:new_streams7} and \ref{fig:new_streams8} show new stream candidates that have been only found in \vma{} and not in \vmc.

\clearpage

\bibliographystyle{aa} 
\bibliography{literature}

\end{document}